\documentclass[ aip, amsmath,amssymb,preprint]{revtex4-1}

\usepackage{graphicx}% Include figure files
\usepackage{dcolumn}% Align table columns on decimal point
\usepackage{bm}% bold math
\usepackage{color}
\usepackage{ulem}
\usepackage{color}

\usepackage{hyperref}
\usepackage{float}
\usepackage{wrapfig}
\usepackage{cancel}
\newcommand{\superscript}[1]{\ensuremath{^{\textrm{#1}}}}
\newcommand{\subscript}[1]{\ensuremath{_{\textrm{#1}}}}

\definecolor{readgreen}{rgb}{0.2,0.45,0.10}

\begin{document}

\title[]{Linear Scaling Density Matrix Real Time TDDFT: Propagator Unitarity \& Matrix Truncation}% Force line breaks with \\

\author{Conn O'Rourke}
 \affiliation{London Centre for Nanotechnology, University College London, 17-19 Gordon St, London, WC1H 0AH, United Kingdom}%Lines break automatically or can be forced with \\
 \altaffiliation[UCL Satellite, MANA]{ International Centre for Materials Nanoarchitectonics (MANA), National Institute for Materials Science (NIMS), 1-1 Namiki, Tsukuba, Ibaraki 305-0044, Japan\\}

 \email{ucapcor@ucl.ac.uk}
\author{David R. Bowler}%
 \affiliation{London Centre for Nanotechnology, University College London, 17-19 Gordon St, London, WC1H 0AH, United Kingdom}%
 \altaffiliation[UCL Satellite, MANA]{ International Centre for Materials Nanoarchitectonics (MANA), National Institute for Materials Science (NIMS), 1-1 Namiki, Tsukuba, Ibaraki 305-0044, Japan}

\date{\today}

\begin{abstract}
Real time, density matrix based, time dependent density functional theory proceeds through the propagation
of the density matrix, as opposed to the Kohn-Sham orbitals.
It is possible to reduce the computational workload by imposing spatial cut-off radii on sparse matrices, 
and the propagation of the density matrix in this manner provides direct access to the optical response of 
very large systems, which would be otherwise impractical to obtain using the standard formulations of TDDFT. 
Following a brief summary of our implementation, along with several benchmark 
tests illustrating the validity of the method, we present an exploration of the 
factors affecting the accuracy of the approach. In particular we investigate the effect of basis set 
size and matrix truncation, 
the key approximation used in achieving linear scaling, on the propagator unitarity and optical spectra.
Finally we illustrate that, with an appropriate density matrix truncation range applied, the computational 
load scales linearly with the system size and discuss the limitations of the approach.
\end{abstract}

%Use showkeys class option if keyword
                              %display desired
\maketitle

\section{\label{Intro}Introduction}

Linear scaling or $\mathcal{O}(N)$ density functional theory (DFT), in which the 
computational workload scales linearly with the number of atoms in the system $N$, 
is now well established \cite{bowler_ON}. 
In the standard approach to DFT, diagonalisation of an eigenvalue equation, or alternatively the 
orthogonalisation of the Kohn-Sham states during minimisation of the energy, 
results in a severe computational bottleneck that limits the size of systems which can be studied.
Working with the density matrix, upon which a truncation radius is applied, allows the
computational workload to be made to scale linearly with $N$. Circumventing the size limitations of 
the standard approach in this manner allows vastly larger systems to be studied: for example calculations
have now been performed on millions of atoms \cite{conquest_million,vdV_million}, in comparison to the upper limit of 
around a thousand for the standard approach.

While density functional theory is a ubiquitous tool in the arsenal of the electronic structure
theorist, it is limited to the study of \textit{ground-state} properties. Extending DFT to the time 
domain results in its excited state couterpart, time dependent density functional theory (TDDFT).
Linear response TDDFT (LR-TDDFT), as developed by Casida\cite{casida1}, again suffers from a computational 
bottleneck which forces it to scale poorly with system size. LR-TDDFT requires the 
solution of an eigenvalue equation for a matrix written in the space of electron-hole pairs, which ostensibly
scales as poorly as $\mathcal{O}(N^{6})$. In practice this scaling can be reduced, through efficient 
implementation and methods employing the Liouville-Lanczos approach, to be as low as
$\mathcal{O}(N^{3})$\cite{parsec,turbo}. For small systems LR-TDDFT is computationally feasible, and has
been widely used, while for larger systems the scaling renders it unsuitable. It is also worth noting that 
linear scaling density matrix based LR-TDDFT, avoiding the propagation of the density matrix, has also been 
recently demonstrated\cite{onetep_linear}.

An alternative approach to LR-TDDFT is the real time propagation of the time-dependent Kohn-Sham
equations, pioneered by Yabana and Bertsch\cite{Yabana}. Real time TDDFT (RT-TDDFT) proceeds by 
the construction of an effective  Hamiltonian, followed by the direct propagation of the Kohn-Sham orbitals
 using this Hamiltonian.
Assuming both the number of occupied states (N\subscript{KS}) and the number of mesh 
points (N\subscript{M}) scale linearly with system size, RT-TDDFT will scale
with the number of atoms, N, as $N_{KS} N_{M} \sim N^{2}$. A significant prefactor in the form of 
the number of time steps and the computational effort for construction of the Hamiltonian exists, 
making this method unsuitable for systems of small size. However the $\mathcal{O}${(N\superscript{2})} 
scaling have made it the natural choice for tackling systems of large size, and a complementary partner 
to Casida's approach.

In a similar manner to $\mathcal{O}(N)$ DFT, it is possible to improve upon the scaling of RT-TDDFT by
propagating the density matrix, as opposed to propagating the Kohn-Sham orbitals directly.
By applying a spatial truncation radius upon the density matrix, the computational workload 
can be reduced, opening up the possibility of studying excited states
in large systems that cannot feasibly be examined with other methods. Although not widely employed, 
this approach has been demonstrated to scale linearly with system size, and has been used
to study several large systems; fullerene, sodium clusters, polyacetylee oligomers, carbon 
nanotubes and silicon clusters to name a few\cite{RDM1,RDM2,RDM3,RDM4,RDM5,RDM6,RDM7}. 

Several factors must be taken into consideration when employing this method, for example the accuracy 
of results produced will depend strongly on the range of truncation of the density matrix.
Also when working in a non-orthogonal basis, as is the case in the CONQUEST code, the overlap matrix will 
be well-ranged. However the inverse overlap, which features in the density matrix propagators, 
will not necessarily be. In order to ensure the unitarity of the propagation the propagtors must be 
carefully tested for matrix truncation errors, and little discussion on the effect of matrix truncation 
upon propagator unitarity have been presented elsewhere. 

In this paper we briefly summarize our implementation of RT-TDDFT in the CONQUEST code, for completeness,
 and confirm
its reliability. We then present several tests probing the limitations of the method, and factors affecting
accuracy. In particular we examine the effect of matrix truncation, the key approximation used in 
achieving linear scaling, on the unitarity of the propagators used and optical spectra generated. 

\section{\label{Comp}Computational Approach}

Linear scaling approaches for excited state properties have existed for well over a decade (for a review see \cite{chen_on_td_rev}), with the 
first approach utilising the locality inherent in the density matrix and being carried out at the 
semi-empirical level\cite{RDM1}. Subsequent efforts again all tend to employ the nearsightedness of the 
density matrix, with the first full linear scaling TDDFT being done by Yam et. al. 
\cite{ON_TDDFT_1,chen_alkane}. Our approach follows that of Yam et. al. closely, with a few differences; most
notably we choose not to perform the orthogonalisation procedure via the Cholesky decomposition and rather
work in our non-orthogonal basis. As mentioned, linear scaling approaches to calculating the excited state 
properties in the frequency domain have also been presented, by Yokojima et. al.\cite{RDM1,RDM3}, and more 
recently by Zuehlsdorff et. al. in the ONETEP code \cite{onetep_linear}. It is also worth noting that an 
approach for calculating the unoccupied Kohn-Sham states, via a basis 
optimisation approach which is also linear scaling, has also been implemented in the ONETEP code\cite{onetep_dos}.  

In a similar vein to the standard approaches to TDDFT in the time and frequency domain, the 
reformulations using the density matrix can be viewed as complementary to one another. The frequency
domain approach is suitable for calculating the lowest optical excitations in the system, but if the 
density matrix response involves higher excitations it will not be suitable. While the
real-time density matrix approach employed here and by Yam et. al. calculates the full optical spectrum, 
it has a significant prefactor in the form of the number of time steps needed for the numerical integration.

In this section we briefly give an overview of the approach in our non-orthogonal basis set, and in the
subsequent section we illustrate the effect of the basis set on the results, and the reliability of the 
method with several tests on small molecules.

\subsection{\label{DM-TDDFT}Density Matrix RT-TDDFT}
Rather than working with the conventional single particle Kohn-Sham
orbitals, CONQUEST works with the density matrix written in a
seperable form in terms of a localised basis of support functions $\phi_{i\alpha}$

\begin{equation}
\rho(\textbf{r},\textbf{r}^{\prime}) = \sum_{i\alpha,j\beta} \phi_{i\alpha}(\textbf{r}) \textit{\textbf{K}}_{i\alpha,j\beta}\phi_{j\beta}(\textbf{r}^{\prime})
\end{equation}

where $\phi_{i\alpha}$ is the $\alpha^{\text{th}}$ support function
centred on atom \textit{i}. Support functions are a non-orthogonal basis
set of localised orbitals, and have an overlap matrix given by:

\begin{equation}
\textit{\textbf{S}}_{\alpha,\beta} 
 =\int \! \phi_{i\alpha}\left(\textbf{r}\right) \phi_{j\beta}(\textbf{r}) \, \mathrm{d} \textit{\textbf{r}}
\end{equation}

Linear scaling behaviour can be obtained through applying a spatial cut-off
on the density matrix. Beyond this cut-off radius the matrix elements are set to zero which,
 along with the spatial limitation of the support functions, ensures that
the number of non-zero density matrix elements increases
linearly with system size (for a fuller overview of the CONQUEST code see \cite{Conquest1}).

RT-TDDFT is now well established\cite{Yabana}, and implementations of 
density matrix RT-TDDFT have been reported elsewhere\cite{chen_on_td_rev,chen_alkane}.
Rather than employing an orthogonalisation procedure via a Cholesky or L\"{o}wdin
decomposition, which will increase the range of the sparse matrices and is done elsewhere, we work in our
non-orthogonal basis.
Expanding the time-dependent Kohn-Sham equations in this basis of
non-orthogonal support functions, in the instance where the
support functions are stationary with time, gives:

\begin{align}
\textit{i}  \frac{\partial}{\partial t} 
\textit{\textbf{c}}(\textit{t})  &= \textit{\textbf{S}}^{-1}\textit{\textbf{H}}
\textit{\textbf{c}}(\textit{t})\\
\text{and}&\nonumber\\
\textit{i}  \frac{\partial}{\partial t} 
\textit{\textbf{c}}^{\dagger}(\textit{t})   & = - 
\textit{\textbf{c}}^{\dagger}(\textit{t})\textit{\textbf{H}}\textit{\textbf{S}}^{-1}
\end{align}

which describe the time dependence of the coefficients of our
basis set expansion, $\textbf{c}(t)$.
This allows us to write the quantum Liouville equation of
motion for our auxiliary density matrix $\textbf{K}$ in the
non-orthogonal support function basis:
\begin{equation}
\textit{i}  \dot{\textit{\textbf{K}}} 
 =\textit{\textbf{S}}^{-1}\textit{\textbf{H}\textbf{K}} - \textit{\textbf{K}\textbf{H}\textbf{S}}^{-1}
\end{equation}

The formal solution to this equation can be expressed as:
\begin{equation}
\textbf{\textit{K}}(\textit{t})=\textit{\textbf{U}}(\textit{t},\textit{t}_{0})\textit{\textbf{K}}(\textit{t}_{0})\textit{\textbf{U}}^\dag(\textit{t}_{0},\textit{t})
\end{equation}
where $\textit{\textbf{U}}(\textit{t,t}_{0})$ is a propagator satisfying both:
\begin{align}
\textbf{\textit{c}(\textit{t})}&=\textbf{\textit{U}}(\textit{t},\textit{t}_{0})\textbf{\textit{c}}(\textit{t}_{0})\\
\textit{i} \frac{\partial}{\partial \textit{t}} 
\textbf{\textit{U}}(\textit{t},\textit{t}_{0})&= \textbf{\textit{S}}^{-1}\textbf{\textit{H}} 
\textbf{\textit{U}}(\textit{t},\textit{t}_{0}) 
\end{align}

Expressing the propagator in integral form we have:
\begin{equation}
\textbf{\textit{U}}(\textit{t},\textit{t}_{0}) = \mathcal T  \exp \left\{-\textit{i}\int_{\textit{t}_{0}}^{\textit{t}}\mathrm{d}\, \tau \textbf{\textit{S}}^{-1}\textbf{\textit{H}}(\tau)\right\} 
\end{equation}
where $\mathcal T$ is the time ordering operator. Evolution of the
system for a total time, $\textit{T}=n\Delta \textit{t}$, may be carried out
piecewise in smaller intervals, allowing us to express the
total evolution operator as the product of small time operators:
\begin{equation}
\textbf{\textit{U}}(t,t_{0}) \simeq \prod_{n=0}^{N-1} \textbf{\textit{U}}\left(\left(n+1\right)\Delta \textit{t},n\Delta \textit{t}\right)
\end{equation}
where
\begin{equation}
\textbf{\textit{U}}(\textit{t}+\Delta \textit{t},\textit{t}) = \exp\left[-i \textbf{\textit{S}}^{-1}\textbf{\textit{H}}(\tau)\Delta \textit{t} \right]
\label{prop}
\end{equation}

Evolution of the time dependent system is then reduced to
the problem of approximating the propagator $\textbf{\textit{U}}(\textit{t}+\Delta \textit{t},\textit{t})$.
Two approximations exist in the definition
of $\textbf{\textit{U}}(\textit{t}+\Delta \textit{t},\textit{t})$, firstly that of approximating
the matrix exponential $\exp(\textbf{A})$ and secondly the
exact form of the matrix for which we wish to calulate the
exponential. There are
several methods for calulating the exponential of a matrix
\cite{matrix_exp}, here we use the simplest approximation,
a Taylor expansion:

\begin{equation}
\exp(\textbf{A} \Delta \textit{t}) = \textbf{I} + \sum_{n=1}^{\infty} \frac{(\textbf{A}\Delta \textit{t})^{n}}{n!}
\label{matrix_exp}
\end{equation}

Similarly there are many different approaches for deciding which
matrix exponential to use as a propagator. Three approximations
have been implemented: the so called \textit{exponential-midpoint} propagator (EM),
the \textit{enforced time-reversal symmetry} (ETRS) propagator and the fourth order
Magnus (M4) propagators, all of which are taken from the work of Marques
et al.\cite{marques_prop} on RT-TDDFT propagators, and are briefly described in our
non-orthogonal basis for completeness.

The exponential midpoint propagator approximates the \textbf{U}$(t+\Delta t,t)$
by the exponential taken at $\tau=t+\Delta t/2$:
\begin{equation}
\textbf{\textit{U}}_{EM}(\textit{t}+\Delta \textit{t},\textit{t}) = \exp \left\{- \textit{i} \textbf{\textit{S}}^{-1}\textbf{\textit{H}}\left(\textit{t}+\frac{\Delta \textit{t}}{2}\right)\right\}
\label{EM}
\end{equation}

Implicitly enforcing time-reversibility, such that propagating forward
from $t$ and backwards from $\textit{t}+\Delta \textit{t}$ by $\Delta \textit{t}/2$ produce the
same result, provides the so called \textit{enforced time-reversal
symmetry} method:
\begin{align}
\textbf{\textit{U}}_{ETRS}(\textit{t}+\Delta \textit{t},\textit{t})=&\exp \left\{ -\textit{i} \frac{\Delta \textit{t}}{2} \textbf{\textit{S}}^{-1}\textbf{\textit{H}}\left(\textit{t}+\Delta \textit{t}\right) \right\} \notag\\ 
&\times\exp\left\{ -\textit{i} \frac{\Delta \textit{t}}{2} \textbf{\textit{S}}^{-1}\textbf{\textit{H}}\left(\textit{t}\right) \right\}
\end{align}

Using the Magnus operator the exponential solution to Schr\"{o}dinger
equation for a time-dependent Hamiltonian may be written as\cite{M4}:
\begin{equation}
\textbf{\textit{U}}_{M4}(\textit{t}+\Delta \textit{t},\textit{t})=\exp \left\{ M_{G4}\right\} 
\end{equation}
where \textbf{M}\subscript{G4} is an infinite series of integrals
providing an exact solution. Truncating this expansion to
fourth order and approximating the integrals using Gauss-Legendre
points as in \cite{marques_prop} gives in our non-orthogonal basis:
\begin{align}
M_{G4}=&-\textit{i} \frac{\Delta \textit{t}}{2}\left[ \textbf{\textit{S}}^{-1}\textbf{\textit{H}}(\textit{t}_{1})+\textbf{\textit{S}}^{-1}\textbf{\textit{H}}(\textit{t}_{2})\right] \notag\\
&- \frac{\sqrt{3}\Delta \textit{t}^{2}}{12} \left[ \textbf{\textit{S}}^{-1}\textbf{\textit{H}}(\textit{t}_{2}),\textbf{\textit{S}}^{-1}\textbf{\textit{H}}(\textit{t}_{1})\right] 
\end{align}
where $\textit{t}_{2,1}=\textit{t}+[1/2 \pm \sqrt{3}/6]\Delta \textit{t}$.

It is important to note the presence of the inverse overlap matrix $\textbf{\textit{S}}^{-1}$ in these
propagators, and again consider that while the overlap matrix will be well-ranged and suitable for 
truncation, the inverse overlap is not necessarily so. We therefore need to carefully test the sparsity of the 
product $\textbf{S}^-1\textbf{H}$, and its effect on the unitarity of our propagators.
%**********************************************************************************
\subsection{\label{LR}Linear Response}

The idea behind extracting optical transitions from the linear response of a system to an external
electric field is well known\cite{Yabana,RDM7}.
Propagating in real time provides direct access to the time-dependent
charge density, and therefore the electronic response to external fields.
Applying a time dependent external electric field polarised along
axis \textit{j},
\begin{equation}
\delta \textit{v}_{ext}\left(\textbf{\textit{r}},\textit{t}\right)=-\textbf{\textit{E}}_{\textit{j}}(\textit{t}) \cdot  \textbf{\textit{r}} \nonumber
\end{equation}
allows us to examine the time-dependent response of the system.
Application of this
electric field will produce an induced time-dependent dipole moment:
\begin{equation}
\textbf{\textit{P}}\left(\textit{t}\right)=\textbf{\textit{P}}\left(0\right)-\int\mathrm{d}\textbf{\textit{r}}\, n\left(\textbf{\textit{r}},\textit{t}\right)\textbf{\textit{r}} . 
\end{equation}

\begin{figure}
  \begin{center}
%   \begin{tabular}{c c}
%     \begin{minipage}[h]{0.3\textwidth}
%      \includegraphics[trim = 0mm 0mm 0mm 0mm, clip, width=1.\textwidth]{IMAGES/RESULTS/}
%      \end{minipage}&
        \begin{minipage}[h]{0.45\textwidth}
        \includegraphics[trim = 0mm 0mm 0mm 0mm, clip, width=1.\textwidth]{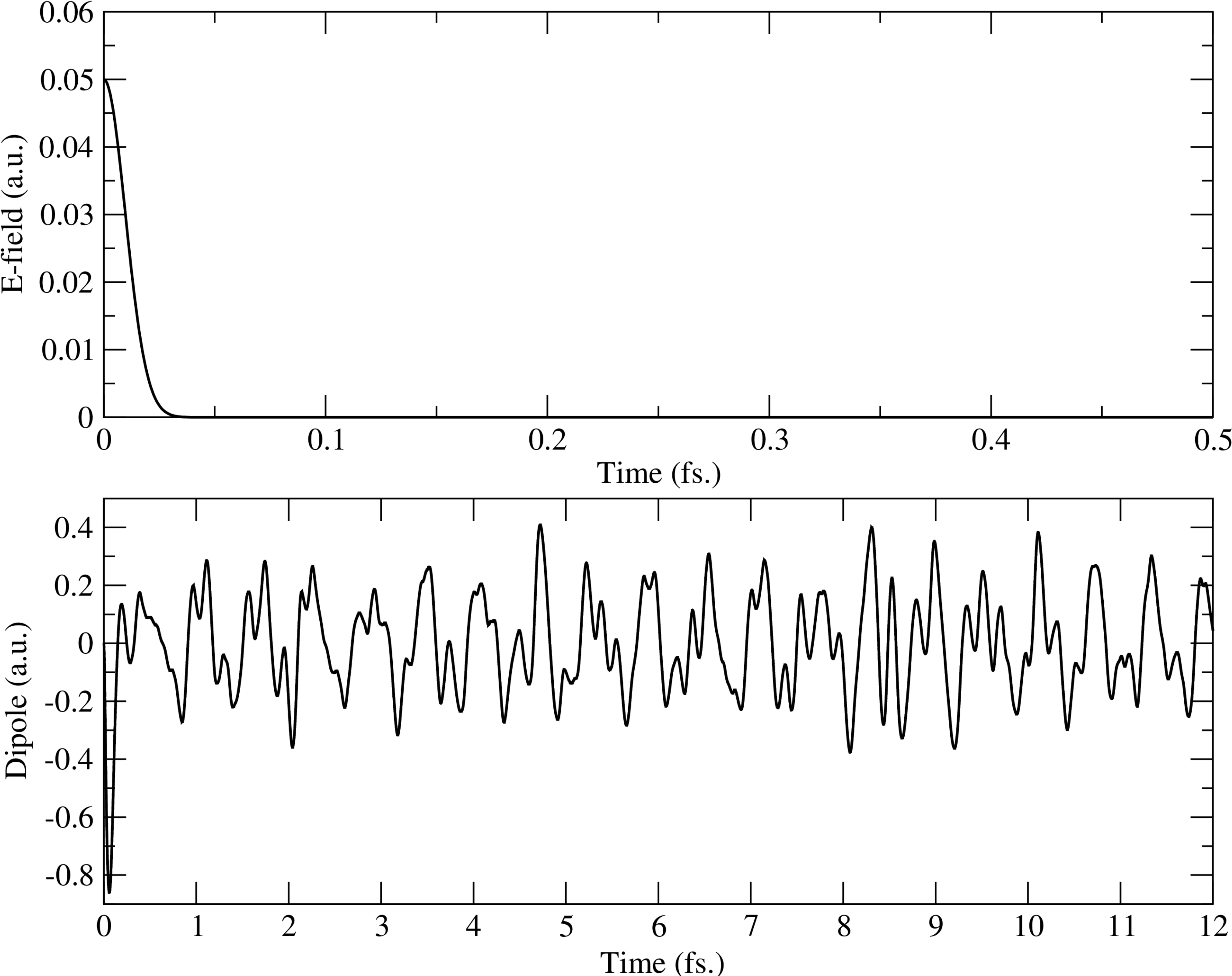}
         \end{minipage}
%        \end{tabular}
     \caption{Applied electric field and induced dipole moment for a
benzene molecule. ($\Delta t = 0.03$ a.u. $\approx$ 0.00073 fs.)}
    \label{benz_dip}
   \end{center}
\end{figure}

As an example of the calculated repsonse of a system to an applied electric
field, figure \ref{benz_dip} illustrates the induced dipole
response of a benzene molecule on application of a field with a
Gaussian time profile, centred at $t=0$.

Access to the time-dependent dipole moment allows us to calculate
the time dependent polarisability:
\begin{equation}
\alpha_{i j}(\omega)=\frac{\int \mathrm{dt}\, e^{i\omega t}\textit{P}_{i}(\textit{t})}{\int \mathrm{dt}\, e^{i \omega \textit{t}} \textit{E}_{j}(\textit{t})} \nonumber
\end{equation}

The imaginary part of the polarisability is directly proportional to the
absorption cross section, $\sigma\left(\omega\right)$ and the
experimentally observed strength function, $S\left(\omega\right)$.

\begin{equation}
%\sigma\left(\omega\right)=\frac{4\pi\omega}{c} Im\left(\frac{1}{3}Tr\left(\alpha_{\mu j}\right)\right)\\
%S\left(\omega\right)=\frac{2 m \omega}{\pi e^{2} \hbar} Im\left(\frac{1}{3}Tr\left(\alpha_{\mu j}\left(\omega\right)\right)\right)
\textit{S}\left(\omega\right)=\frac{2 \omega}{\pi} Im\left(\frac{1}{3}Tr\left(\alpha_{\mu j}\left(\omega\right)\right)\right)
\end{equation}

As noted by Tsolakidis et. al., the approach satisifies the f-sum rule and the integration of the strength function over energy gives the number of electrons, which may be used as a measure of the completeness of the
basis set\cite{RDM7}.

Density matrix RT-TDDFT therefore has the potential to be an extremely
useful tool for theoretically predicting the electronic absorption
spectra of large system.

\section{\label{Molecules}Small Molecules}
\begin{center}
\begin{table*}[t]
%\begin{center}
\begin{minipage}{1.\textwidth}
\begin{center}
\begin{tabular}{c c c c c c c c c c c}
\hline
\hline

Transition & \multicolumn{8}{c}{Basis Set}& Ref\cite{ethylene_tddft}& Expt. \cite{ethylene_expt}\\
           & 2ZP& 2Z2P& 3ZP& 3Z2P& 4ZP& 4Z2P& 5ZP& 5Z2P & 3ZP& \\
\hline
\hline
1 meV&&&&&&&&&\\
$\pi \rightarrow \pi^{*}$& 7.84& 7.62& 7.73& 7.62& 7.67& 7.62& 7.67& 7.62 & 7.45& 8.0\\
$\pi \rightarrow 3s$& 8.43& 7.95& 7.78&7.67& 7.46& 7.40& 7.46&7.29 & 6.69& 7.11\\
\hline
5 meV&&&&&&&&&&\\
$\pi \rightarrow \pi^{*}$& 7.82& 7.73& 7.75& 7.69& 7.70& 7.68& 7.67&7.67& 7.45& 8.0\\
$\pi \rightarrow 3s$& 10.64& 8.03& 7.88& 7.76& 7.57& 7.51& 7.46& 7.45&
6.69& 7.11\\

\hline
\hline
\end{tabular}
\end{center}
\end{minipage}
%\normalsize
\caption{Basis set dependence of calculated TDLDA transition energies (eV.) for first valence ($\pi \rightarrow \pi^{*}$) and Rydberg ($\pi \rightarrow 3s $) excitations for the C\subscript{2}H\subscript{4} molecule.}
%\end{center}
\label{basis_size}
\end{table*}
\end{center}

In order to verify that our implementation is correct we have
performed tests on several systems for which the electronic
transitions have been studied experimentally and theoretically
elsewhere, allowing us to make direct comparisons.
For this purpose we have chosen four
small molecules (Carbon monoxide, Methane, Ethylene and Benzene) and
used our implementation to calculate the optical absorption
spectra within the TDLDA approximation.

%\subsubsection{Identifying Transitions}
Meaningful comparison of our results with experiment requires the identification
the electronic transitions to which the peaks in our calculated absorption
spectra correspond. As we have mentioned in Casida's approach information about electronic transitions is
inherently produced, while in RT-TDDFT it is not.

It is often possible to identify the corresponding transition
by examining the polarisation and energy of peaks and comparing to
that of optically allowed transitions experimentally.
Where possible, in order to more confidently assign peaks of our calculated
absorption spectra to particular electronic transitions, we have followed the
procedure in \cite{ethylene_assignment} whereby a sinusoidal electric field tuned
to a particular excitation mode is applied. A resulting electronic resonance is set
up, allowing us to examine the difference between ground state charge
density and excited state charge density and thereby infer the electronic
transition.\\ 

\subsection{Basis Sets}
\begin{table*}[t]
\begin{minipage}{.75\textwidth}
\begin{center}
\begin{tabular}{m{0.12\textwidth} m{0.14\textwidth} m{0.17\textwidth} m{0.12\textwidth} m{0.1\textwidth}}
\hline
\hline
Molecule &  Transition  &  RT-TDDFT  & Ref  & Expt\\
         &              &   (eV.)    &   &     \\
\hline
\hline
CO   &  $\sigma \rightarrow \pi^{*}$ & 8.17 & 8.20\cite{chunping} & 8.51\cite{CO_expt}\\
CH\subscript{4} & T\subscript{2} $\rightarrow$ 3s & 9.22 & 9.27\cite{tddft-ref} & 9.70\cite{methane_exp}\\
C\subscript{2}H\subscript{4}& $\pi \rightarrow \pi^{*}$ & 7.48 & 7.45 \cite{tddft-ref}& 8.00\cite{ethylene_expt}\\
C\subscript{6}H\subscript{6}& $\pi \rightarrow \pi^{*}$ & 6.87 &$\sim$6.90 \cite{Yab-benz}& 6.90 \cite{Koch_benz}\\
\hline
\hline
\end{tabular}
\end{center}
\end{minipage}
\caption{Comparison of calculated TDLDA transition energies for small molecules with other values and experiment. Conquest results obtained with 5Z4P basis sets, with the exception of benzene (2Z2P). }
%\end{center}
\label{1st-trans}
%\end{center}
\end{table*}

Our support functions are expanded in a basis of
numerical orbitals, in this case pseudo-atomic orbitals
generated following the approach of the Siesta code \cite{siesta1}. These PAOs are
eigenfunctions of the atomic pseudopotentials with a
confinement energy shift used to determine a radial cut-off
for the orbitals, beyond which they are zero. This confinement
energy  provides a single parameter to define the cut
off radii for different orbitals, and is the energy each orbital
obtains on being confined by an infinite potential to
a particular radius. It is clear that a minimal basis with which
ground state properties are accurately reproduced will generally
not be satisfactory for calculating excited state properties, and
therefore we illustrate the basis set dependence of two selected
transitions for the C\subscript{2}H\subscript{4} molecule.

Multiple orbitals per angular momentum channel can be
used (multiple-$\zeta$), with the shape of multiple orbitals
determined by a split norm procedure\cite{siesta1}. This procedure
uses a parameter to define the norm of a numerical orbital outside
some radius where they match the tail of the first zeta PAO, and
within this radius the vary smoothly to the origin.
%as $r^{l}(a-br^{2})$\cite{siesta1}.
Subtracting this numerical orbital
from the original PAO gives the multiple-zeta orbital. Of course
it is possible to define these radii by hand and fine tune the basis
set. In addition to multiple zeta, polarisation orbitals can
be included within the basis set, and are obtained by solving
the same pseudo-atomic problem but with an applied electric field.
 
We use the notation SZ, 2Z, 3Z, 4Z to describe single zeta, double zeta,
triple zeta and so on. Similarly we describe the number of polarisation
orbitals included in the basis by SZP, SZ2P and SZ3P
(one, two and three polarised orbitals respectively).

To first gauge the effect of varying our basis set on the results
we have performed calculations on the ethylene molecule with
varying numbers of PAOs and two different confinement energies.
The basis sets have been generated with a confinement energy
of 1 meV and 5 meV, resulting in confinement radii
of 4.93 and 4.24 \AA\ for the carbon atoms respectively,
and radii of 4.77 and 4.21 \AA\ for the hydrogen atoms respectively.
The total run time was
14.51 fs. (600 au.)  with a  time step of
$\sim$0.0242 fs (0.1 au). The results can be seen in
table \ref{basis_size}.

Calculated energies for the
$\pi \rightarrow 3s$ transition show a wide variation with
basis set choice, while the $\pi \rightarrow \pi^{*}$ valence
transition varies less. This is in line with expectation,
given the more diffuse nature of the Rydberg transition we
would expect its description to require  larger basis.
The effect of systematically increasing the number of basis
functions is to improve our results with respect to that of
the reference values. Similarly increasing the cut-off radii,
by reducing the confinement energy, tends to improve the
quality of the result. This is to be expected,
as increasing the size of our basis set, while systematically
increasing the range, will maximise the
variational degrees of freedom available to describe our time
dependent density matrix.

However our values are still far from those computed elsewhere,
and we find generally that for small molecules
it is essential to use a large basis with multiple extended
polarisation orbitals in order to produce results in line with
other works. In addition we find that fine tuning the radial
cut-offs by hand, as opposed to using the confinement energy and split
norm procedure, can allow us to improve the
quality of our results for small molecules.

\subsection{Small Molecule Results}
\begin{figure}[!]
\begin{tabular}{c c}
   \begin{minipage}[t]{0.5\textwidth}
 \includegraphics[trim = 0mm 0mm 0mm 0mm, clip, width=.9\textwidth]{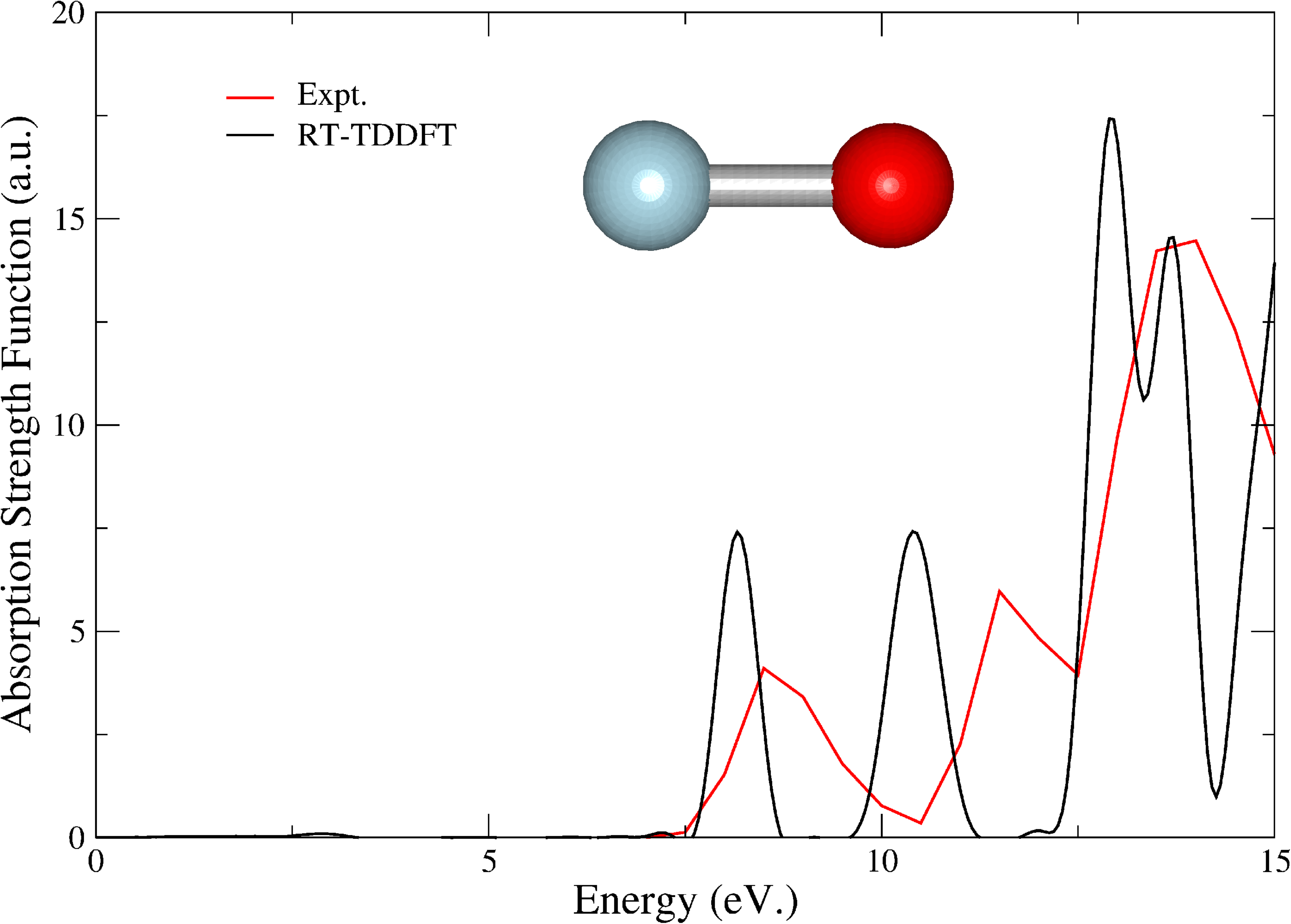}
  \end{minipage}
\\
\textbf{(i)}\\

   \begin{minipage}[t]{0.5\textwidth}
 \includegraphics[trim = 0mm 0mm 0mm 0mm, clip, width=.9\textwidth]{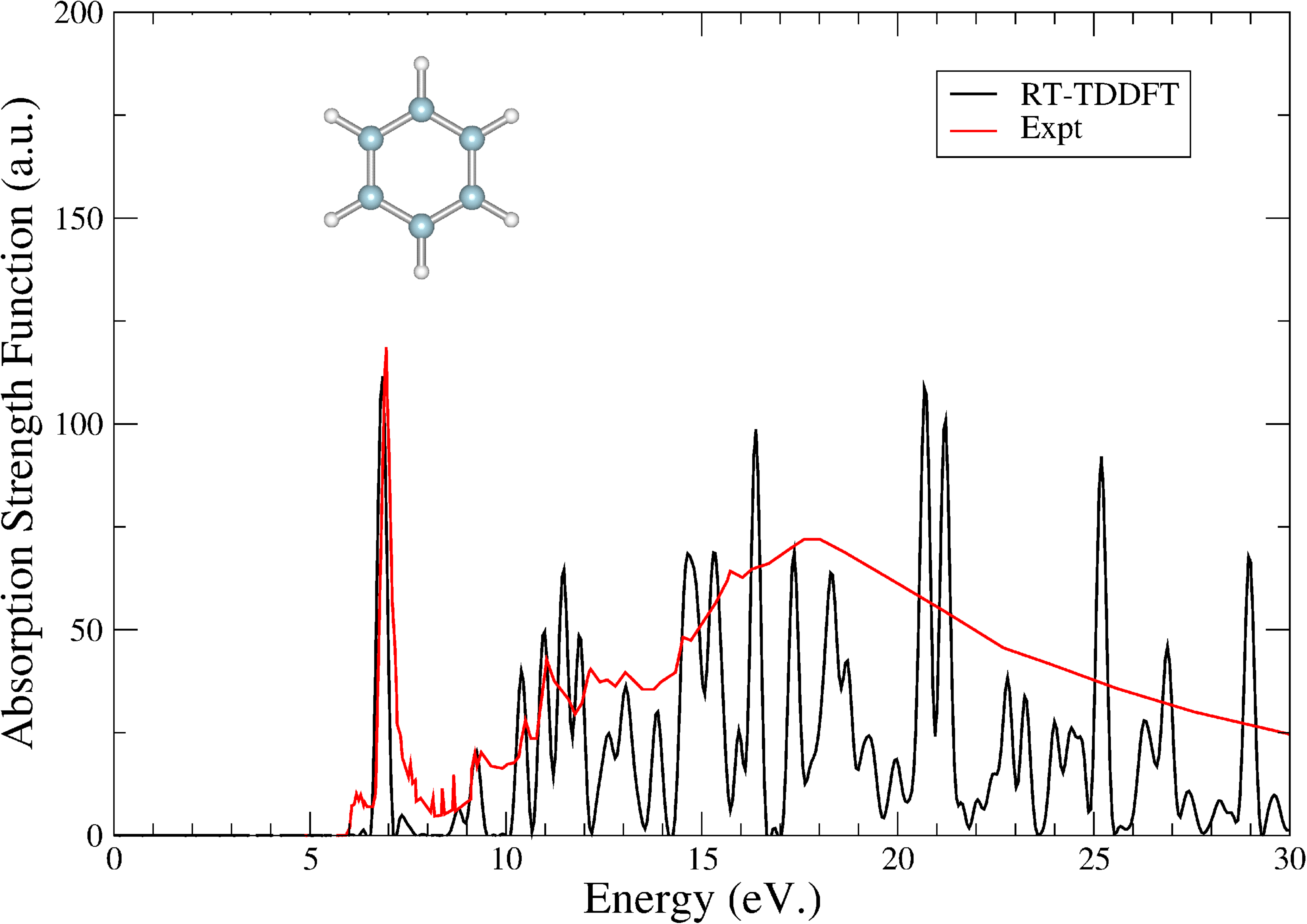}
  \end{minipage}\\
\textbf{(ii)}\\
\end{tabular}
\caption{\textbf{(i)}: Absorption strength function for carbon monoxide from RT-TDDFT and experiment. Experimental data taken from \cite{CO_expt}. \textbf{(ii)}
Absorption strength function for Benzene from RT-TDDFT. Experimental data taken from \cite{Koch_benz}.}
 \label{co_benz_spect}
\end{figure}

Exhibited in table \ref{1st-trans} are the calculated
transitions for our four test molecules. In the case of the
smallest molecules (carbon monoxide, ethylene,
and methane) a hand tuned 5Z4P basis set is employed,
while for benzene the result is obtained using a 2Z2P basis with
a 5meV confinement energy (all the calculations satisfy the f-sum rule to $>94\%$). Also presented in figure \ref{co_benz_spect} are the optical absorption spectra for the benzene and carbon-monoxide molecules,
along with the experimental data.

We can see a strong agreement between our results and that of
other studies, giving us confidence in our implementation.
Very good agreement is exhibited between the calculated
benzene absorption spectra and the experimental values using a
reasonably modest 2Z2P basis set. This highlights
the point that for larger molecules we have generally found
that the need for large hand tuned basis sets, as is necessary
for the smaller molecules, is reduced. Typically results
in agreement with those in the literature and experiment are found
using smaller basis sets, a point that is important to
bear in mind, given the context of linear scaling methods.

\section{Propagator Unitarity}

Having demonstrated the correctness of our implementation and explored the
influence of basis sets, we now turn to our main concern,
the effects of localisation in linear scaling methods on
the accuracy of results.

We wish the total charge in our system to remain stable,
and in order for this to be the case the propagators must be
unitary with respect to the non-orthogonal basis set: 

%\cancel{\begin{align}
% \textit{P}(\textit{t}+\Delta \textit{t})&= \textbf{\textit{c}}^{\dagger}(\textit{t}+\Delta \textit{t})\textbf{\textit{S}}\textbf{\textit{c}}(\textit{t}+\Delta \textit{t})\\
%             &= \textbf{\textit{c}}^{\dagger}(\textit{t})\textbf{\textit{U}}^{\dagger}\textbf{\textit{S}}\textbf{\textit{U}}\textbf{\textit{c}}(\textit{t})\\
%&=\textit{\textit{P}}(t)
%\end{align}}

\begin{equation}
\textbf{\textit{U}}^{\dagger}\textbf{\textit{U}}-\textbf{\textit{I}}=0
\label{s_unit}
\end{equation}
where $\textbf{\textit{U}}$ is our propagator matrix and $\textbf{\textit{I}}$ is the
identity matrix.

From our approximation for the matrix exponential, eq. \ref{matrix_exp},
it can be shown that, if it were exact, our propagators would indeed exhibit
this property. However, as it is impossible for us to store an infinite
sum on our computer, we must truncate our Taylor expansion at some point.
Doing so will introduce errors, with two factors affecting
the scale of the break from unitarity; the time step and the number
of terms in our summation. While we can extend our expansion arbitrarily,
and reduce the time step arbitrarily, we wish to avoid excess computational
 expense by keeping the expansion as small as possible and the time
step as large as possible within some acceptable margin of accuracy.
We can directly examine the unitarity of our propagators through
equation \ref{s_unit}.

\subsection{Time-Step Dependence}
As a test we have examined the extent of the break from unitarity for
a range of time-steps and number of terms in the matrix exponential
expansion. We have used a small molecule for the purpose, benzene, with a small applied electric field perturbation with a Gaussian
profile centered on $t=0$. %(as figure \ref{benz_dip}). 

\begin{figure}
  \begin{center}
    \begin{minipage}[h]{0.5\textwidth}
      \includegraphics[trim = 0mm 0mm 3mm 4mm, clip, width=1.\textwidth]{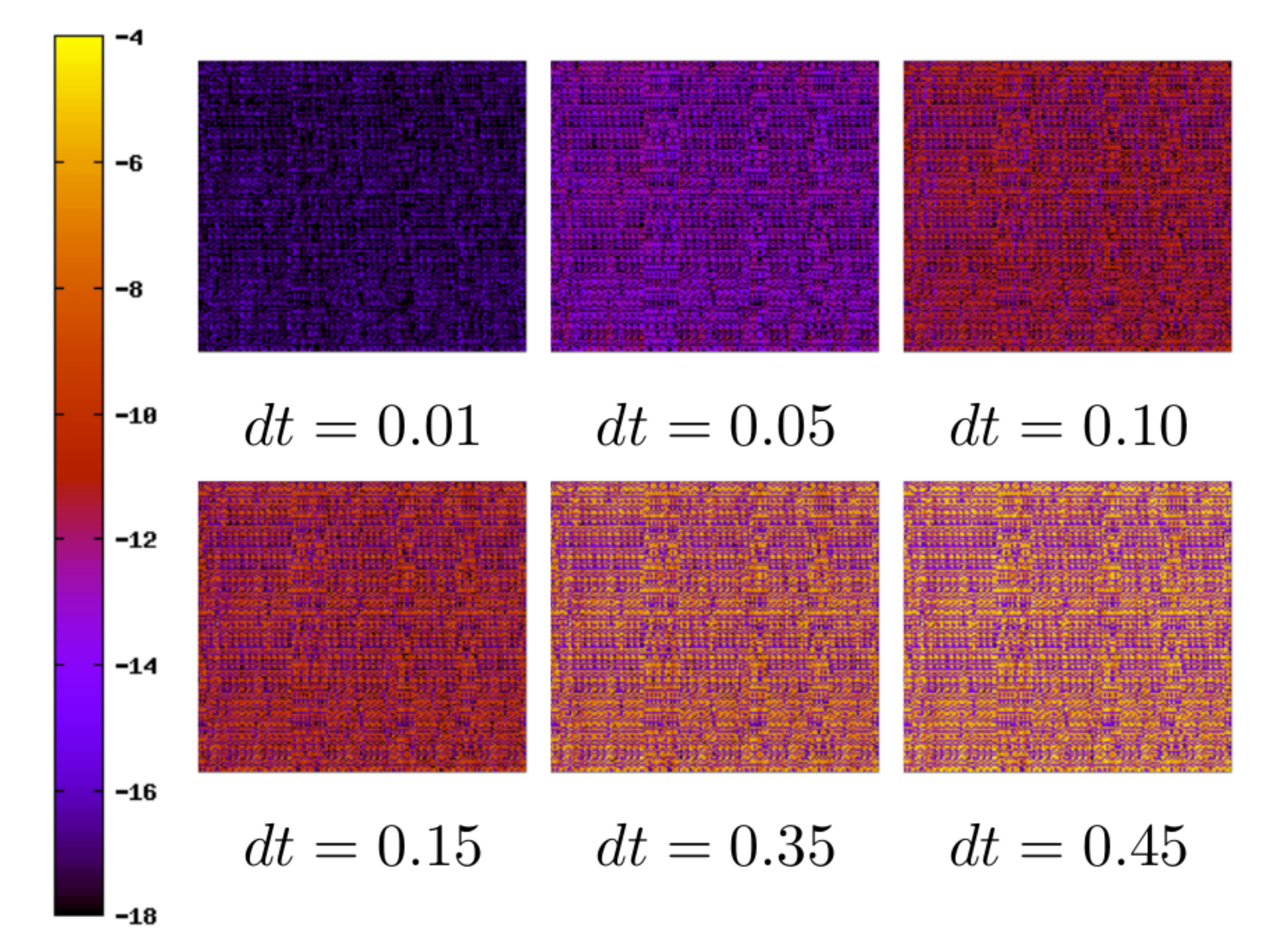}
    \end{minipage}
     \caption{Plot of the absolute values of matrix $\textbf{\textit{U}}^{\dagger}\textbf{\textit{U}}-\textbf{\textit{I}}$ (on a base 10 log scale), illustrating the propagator unitarity for the exponential midpoint propagator, for varying time step sizes (in a.u.). The system studied is a single benzene molecule, and the matrix is shown at the end of a 10 a.u. run. }
    \label{dt_unit}
   \end{center}
\end{figure}

\begin{figure}[!]
   \begin{minipage}{0.49\textwidth}
\rotatebox[origin=c]{0}{\includegraphics[angle=-90,origin=c,trim = 20mm 0mm 20mm 30mm, clip,width=1.\textwidth]{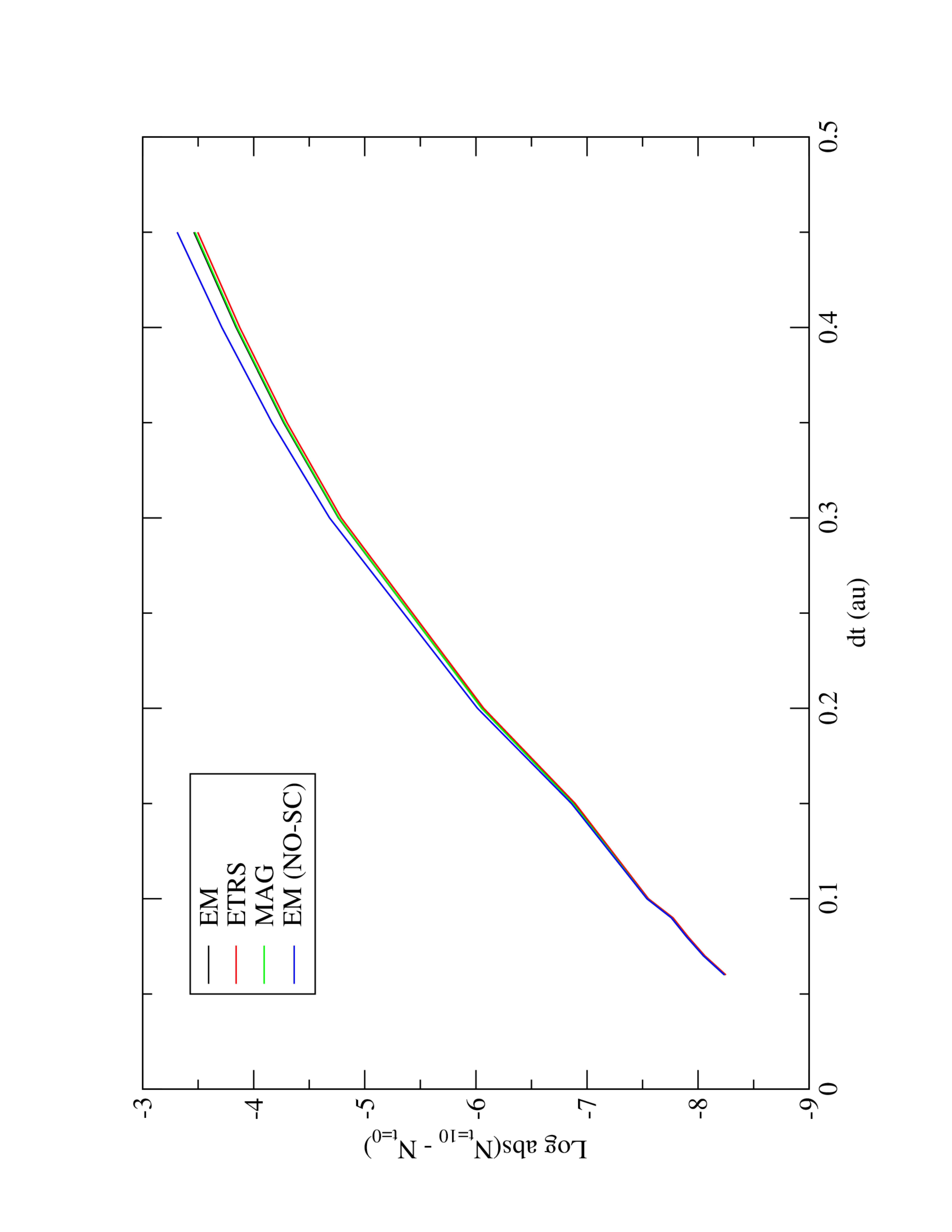}}
  \end{minipage}
\caption{Variation in total charge (on a base 10 log scale) with time step size, following a 10 au. run for benzene using all three propagators. Also included is charge variation for the EM propagator without the self-consistent propagator step (see text for details).}
 \label{dt_chg}
\end{figure}

Exhibited in figure \ref{dt_unit} we can see the dependence on
simulation time step of the propagator unitarity, with the obvious
trend being that as the time step is reduced the propagator
approaches unitarity. We can see that even for time steps up to $\sim 0.15$ a.u. the propagator maintains its unitarity to a high degree
(similar results were obtained for each of the propagators).
The corresponding effect
on the charge conservation can be seen in figure \ref{dt_chg} and, as expected we see that as the time step increases the
conservation of charge deteriorates with the propagation
eventually becoming unstable for large timesteps. While the
maximum permissible timestep will depend on the system under study,
we found that generally a timestep of 0.06 a.u. or below provided
satisfactory charge conservation.

The form of our propagators requires
the extrapolation of the Hamiltonian matrix to
some unknown point beyond the current time t, $\mathbf{H}_{+}$.
As suggested by Marques et al.\cite{marques_prop}
in order to minimise errors it is
possible to carry this procedure out self-consistently. In our case
meaning that we propagate \textbf{\textit{K}}$(t)$ to
\textbf{\textit{K}}$(\textit{t} + \Delta \textit{t})$ based on an extrapolated
Hamiltonian. We then construct a new Hamiltonian
matrix \textbf{\textit{H}}$(\textit{t} + \Delta \textit{t})$ using \textbf{\textit{K}}$(\textit{t} + \Delta \textit{t})$.
$\mathbf{\textit{H}}_{+}$ can then be
interpolated from Hamiltonian matrices for times up
to and including $(\textit{t} + \Delta \textit{t})$, and the whole procedure is
iterated until some self-consistency criteria is obtained.
Generally speaking this procedure is performed three times in the early
stages of a run, following a perturbation, and reduces to two
as the run progresses. The effect of not performing this
self-consistency procedure on the
charge conservation can be seen in figure \ref{dt_chg}.
While the self-consistency cycle is found to improve the
charge conservation, in reality for small time steps the difference
in charge conservation and calculated properties is not
found to be significant enough to warrant the extra computational
load of constructing the Hamiltonian matrix several times
per time-step. As a compromise we enforce the self-consistency
only for a small number of steps ($\sim 50-100$) at the beginning
of a run, typically when our external electric field is
 applied for the study of the linear response and the external
perturbation is largest.

A significant point to note is that little difference is exhibited
between the calculated results using each of the three propagators
in terms of charge conservation, and in general we have found this to
be the case. It
is reported that for systems with strongly time-dependent
Hamiltonians the fourth order Magnus propagator, \textbf{\textit{U}}\subscript{M4},
is advantageous\cite{marques_prop}, but for our present
work this is not the case and we have opted for the simplest
exponential midpoint propagator throughout.

\subsection{Matrix Exponential Truncation}
\begin{figure}{!}
   \begin{minipage}{.49\textwidth}
\rotatebox[origin=c]{0}{\includegraphics[angle=-90,origin=c,trim = 20mm 0mm 20mm 30mm, clip,width=1.\textwidth]{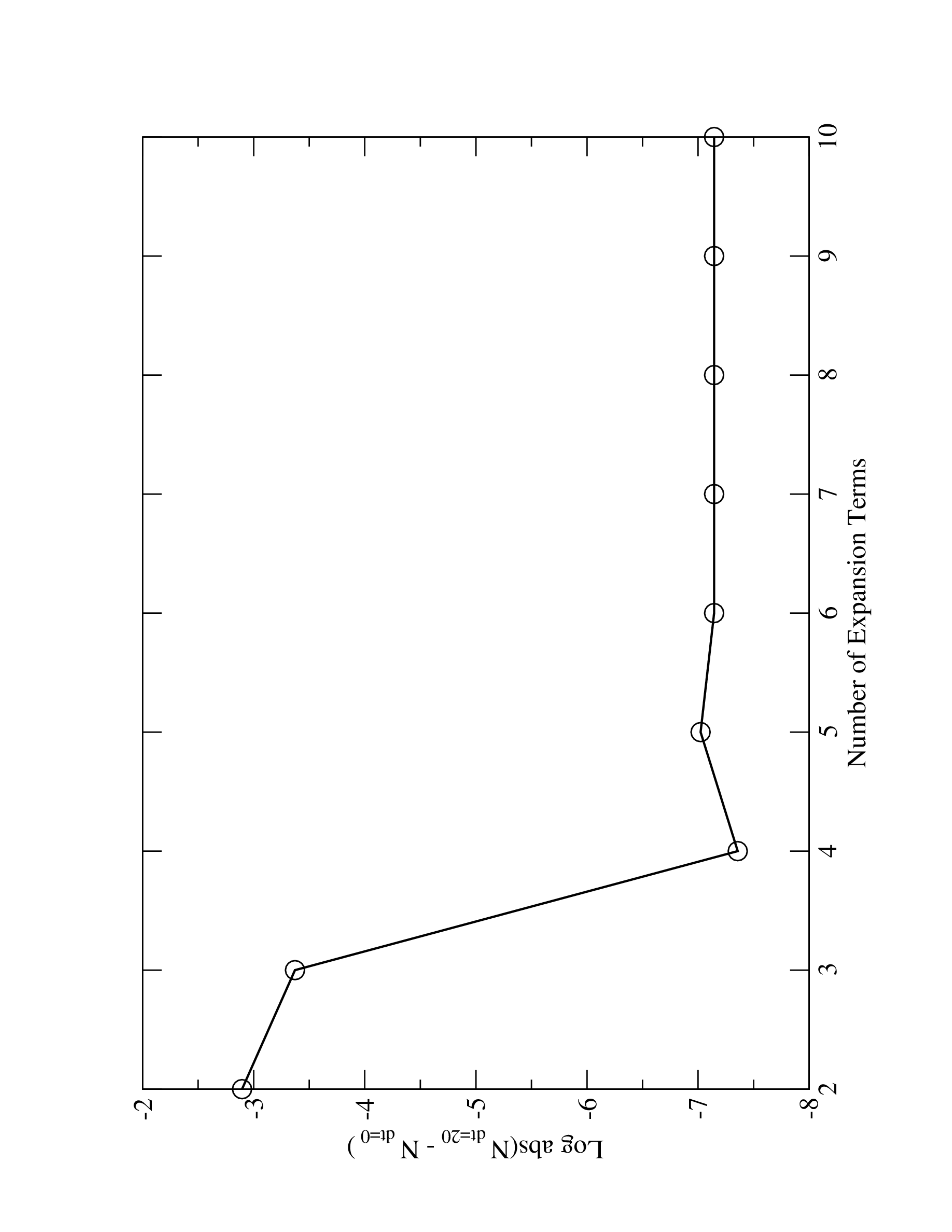}}
  \end{minipage}
\caption{Absolute variation in total charge (on a base 10 log scale) with the number of terms in our matrix exponential expansion, following a 20 au. run for benzene using the EM propagator with a time step of 0.04 au.}
 \label{exp_chg}
\end{figure}

\begin{figure}[!]
  \begin{center}
    \begin{minipage}[h]{0.5\textwidth}
      \includegraphics[trim = 0mm 0mm 3mm 4mm, clip, width=1.\textwidth]{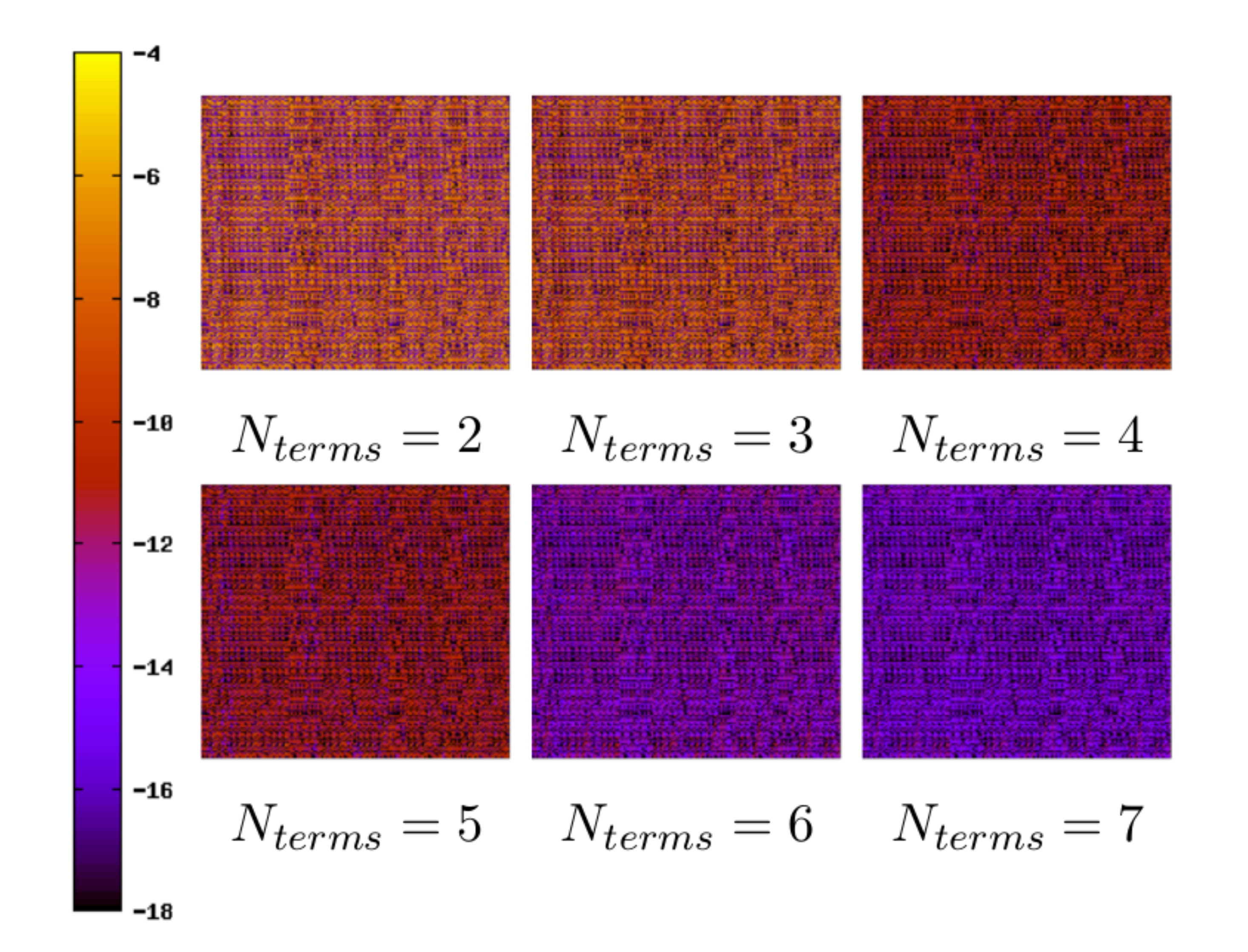}
    \end{minipage}
    \caption{Plot of the absolute values of the matrix $\textbf{\textit{U}}^{\dagger}\textbf{\textit{U}}-\textbf{\textit{I}}$ (on a base 10 log scale), illustrating the propagator unitarity for the exponential midpoint propagator, for differing number of terms in the Taylor expansion for our propagator. The system studied is a single benzene molecule, and the matrix is shown at the end of a 10 au. run (dt = 0.04 au.) }
    \label{exp_unit}
   \end{center}
\end{figure}

The effect of truncating the Taylor expansion used to evaluate
the matrix exponential on the unitarity of the propagator
can be seen in figure \ref{exp_unit}. We see that
reducing the number of terms reduces the unitarity of the
 propagator, as expected.
Looking at figure \ref{exp_chg} the convergence of the charge
conservation with the number of terms in the exponential expansion
can be seen. We find that we reach good convergence with
six terms included in the expansion, and we opt for this
level of accuracy throughout the remainder of the paper.

\section{Alkane Molecules: Testing Matrix Truncation Effects}

In this section we perform calculations on long chain
alkane molecules.%, an example of which can be seen in
%figure \ref{alkane} (C\subscript{11}H\subscript{24})
%along with the general chemical structure for an alkane chain.
Our aim is to examine
the effect of matrix truncation on the propagation of the
density matrix and propagator unitarity, along with the computational
scaling with system size.

%\begin{figure*}
%  \begin{center}
%   \begin{tabular}{c c}
%     \begin{minipage}[h]{0.25\textwidth}
%      \includegraphics[trim = 0mm 0mm 0mm 0mm, clip, width=1.\textwidth]{IMAGES/RESULTS/alk/alk.pdf}
%      \end{minipage}&
%        \begin{minipage}[h]{0.5\textwidth}
%        \includegraphics[trim = 0mm 0mm 0mm 0mm, clip, width=1.\textwidth]{IMAGES/RESULTS/alk/alkane_35.pdf}
%         \end{minipage}\\
%        \end{tabular}
%     \caption{Alkane molecule chemical structure (left) and the molecular structure of the C\subscript{11}H\subscript{24} molecule (right).}
%    \label{alkane}
%   \end{center}
%\end{figure*}

\begin{figure*}
  \begin{center}
   \begin{tabular}{c c}
    \begin{minipage}[h]{0.5\textwidth}
     \includegraphics[angle=-90,origin=c,trim = 90mm 55mm 55mm 95mm, clip, width=1.\textwidth]{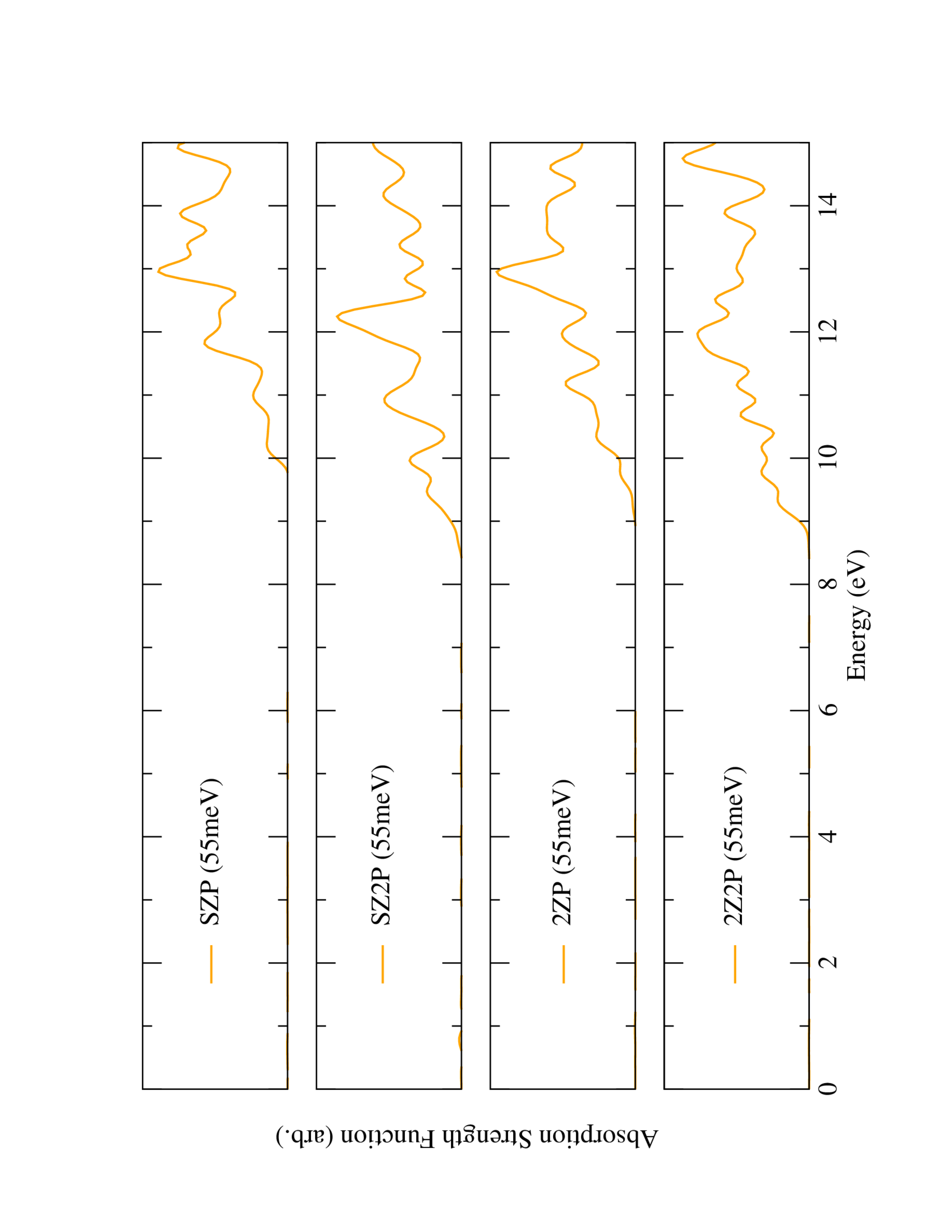}
      \end{minipage}
    &
     \begin{minipage}[h]{0.5\textwidth}
     \includegraphics[angle=-90,origin=c,trim = 90mm 55mm 55mm 90mm, clip, width=1.\textwidth]{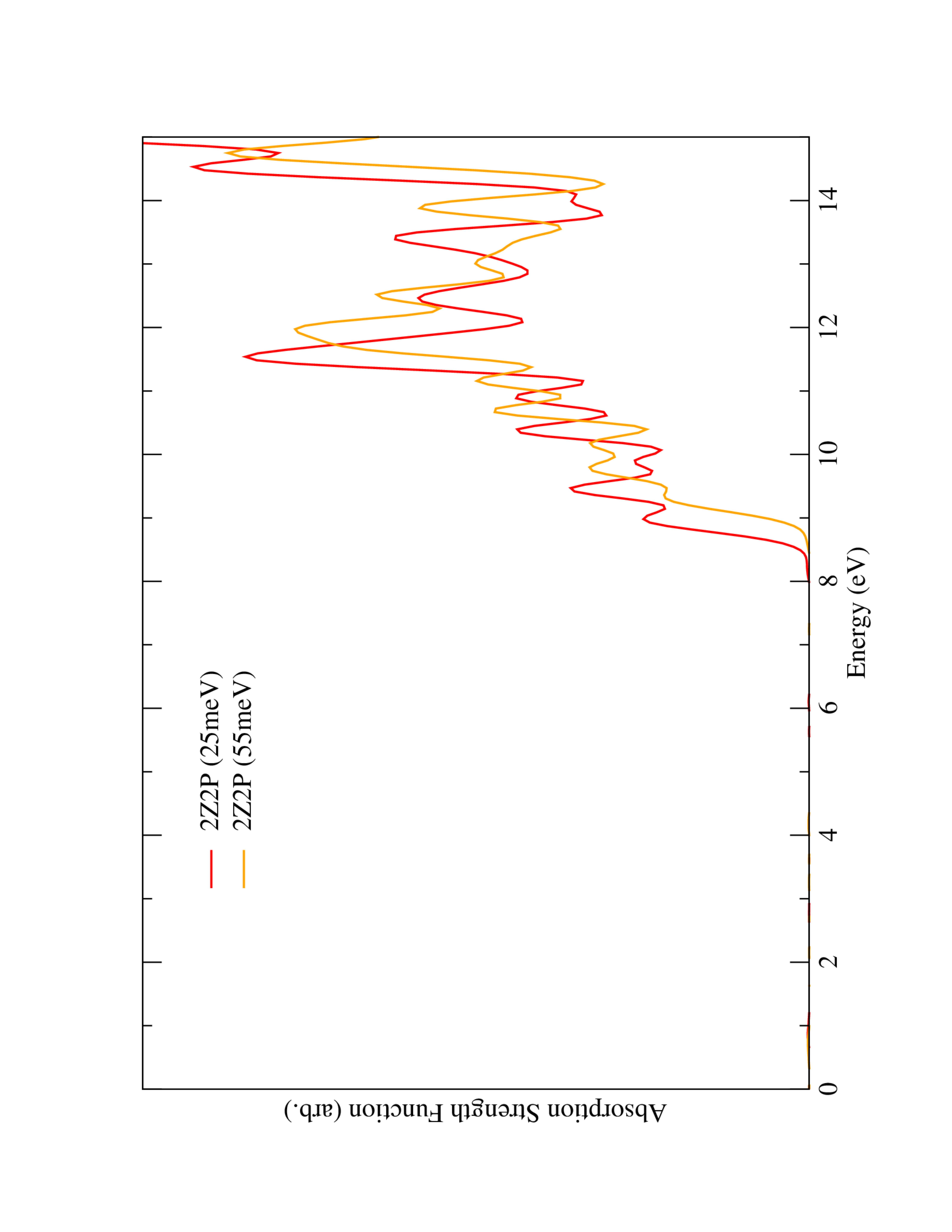}
      \end{minipage}\\
    \textbf{(i)}&\textbf{(ii)}
    \end{tabular}
     \caption{Basis set variation of the calculated alkane optical
absorption spectra. \textbf{(i)}Effect of increasing the number of PAOs in the
basis set and \textbf{(ii)} the effect of extending the radii of the basis functions are shown. }
    \label{alkane2}
   \end{center}
\end{figure*}

As a first step we calculate the absorption spectra for the
C\subscript{11}H\subscript{24} molecule for several different basis
sets using the generalised gradient
PBE functional\cite{pbe} (all further
calculations in this section are performed with this functional), and
 the results can be seen in figure \ref{alkane2}. Experimentally as the
length of the alkane carbon chain increases, the
absorption onset is found to reduce, and the reported adsorption onset
for C\subscript{10}H\subscript{22} is $\sim$ 175 nm. \cite{alkane_expt}
($\sim$ 7.1 eV). We see that as the number of PAOs in the basis set
is increased the calculated absorption onset approaches this value.
Particularly noticeable is the
change of the absorption energy caused by the addition of
polarisation orbitals. Similarly a significant shift is induced
by extending the range of the PAOs (a variation from 55 meV to 25 meV in the confinement energy extends the radii of the carbon and hydrogen basis sets by $\sim$ 0.35 \AA\ and 0.33 \AA\ respectively).
This is understandable, given that the first transitions in the
alkane molecules are reported as being Rydberg in character\cite{alkane_expt}, we would expect the addition of more diffuse PAOs to
improve the description of these excitations.
Given the well documented difficulties of TDDFT to accurately describe
Rydberg transitions \cite{rydberg}, and given that this is not
our aim in any case, we proceed to carry out our tests with the
SZP and SZ2P basis sets generated using a confinement energy of 55meV
(radial cut off for the PAOs is 3.31\AA\ and 3.12\AA\ for
carbon and hydrogen respectively).

Yam et al. have previously studied the long chain alkanes
within the linear scaling excited state
regime \cite{chen_alkane}, calculating the absorption onset
at around 8 eV for C\subscript{40}H\subscript{82} with the LDA
functional. However little discussion of the effects of matrix 
truncation on propagator unitarity have been presented elsewhere.

\subsection{Propagator Truncation}

The use of a basis of non-orthogonal atomic orbitals
requires the inverse overlap matrix for our propagation (indeed
this matrix is required for ground state calculations in any case),
as seen in equation \ref{prop}.
In order to compute the inverse overlap matrix Conquest uses
Hotelling's method\cite{hotelling}, however for poorly conditioned
overlap matrices computing the inverse overlap matrix can prove
difficult. In our current implementation
of TDDFT the atoms remain stationary and so too, therefore,
does the overlap matrix.
Therefore we have also included the possibility of computing the
inverse overlap with the SCALAPACK routines. Although the scaling will
not be linear, computing the inverse overlap
in this way makes only a relatively small contribution to our total
TDDFT runtime,
as we only calculate the inverse overlap once at $t=0$.

\begin{figure}
  \begin{center}
%   \begin{tabular}{c c}
     \begin{minipage}[h]{0.49\textwidth}
     \includegraphics[angle=-90,origin=c,trim = 80mm 25mm 45mm 80mm, clip, width=1.\textwidth]{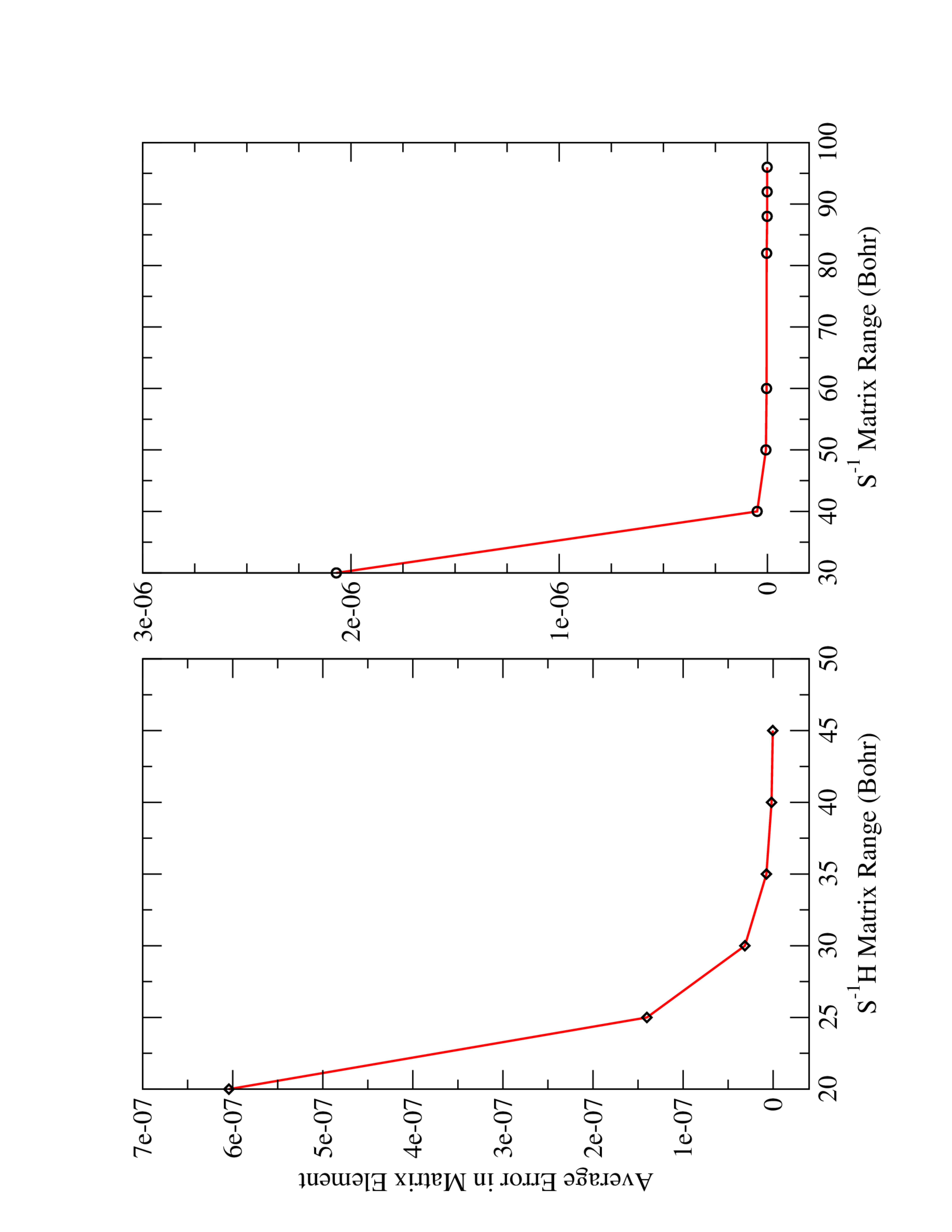}
      \end{minipage}
%    \end{tabular}
     \caption{Average absolute error in the \textbf{\textit{S}}\superscript{-1}\textbf{\textit{H}} (left) and \textbf{\textit{S}}\superscript{-1} (right) matrix elements with matrix range for the C\subscript{47}H\subscript{96} molecule. SZP basis set is used, generated with a 55meV confinement potential.}
    \label{TH_T}
   \end{center}
\end{figure}

While it is apparent that the overlap matrix will be sparse, allowing
it to be truncated, the inverse of a sparse matrix will not
in general be sparse itself. We have therefore tested the
effect of truncating both the \textbf{\textit{S}}\superscript{-1}
matrix and the \textbf{\textit{S}}\superscript{-1}\textbf{\textit{}H} matrix
on the propagation. Figure \ref{TH_T} shows the average
absolute error in the matrix elements of \textbf{\textit{S}}\superscript{-1}
and the \textbf{\textit{S}}\superscript{-1}\textbf{\textit{H}} matrices
caused by truncation (the error in \textbf{\textit{S}}\superscript{-1}
elements given is the average of the elements of the \textbf{\textit{S}}\superscript{-1}\textbf{\textit{S}}-\textbf{I} matrix, and the error in the \textbf{\textit{S}}\superscript{-1}\textbf{\textit{H}} is calculated with the values from an untruncated
 \textbf{\textit{S}}\superscript{-1} matrix).

\begin{figure}
  \begin{center}
    \begin{minipage}[h]{0.5\textwidth}
      \includegraphics[trim = 0mm 0mm 3mm 4mm, clip, width=1.\textwidth]{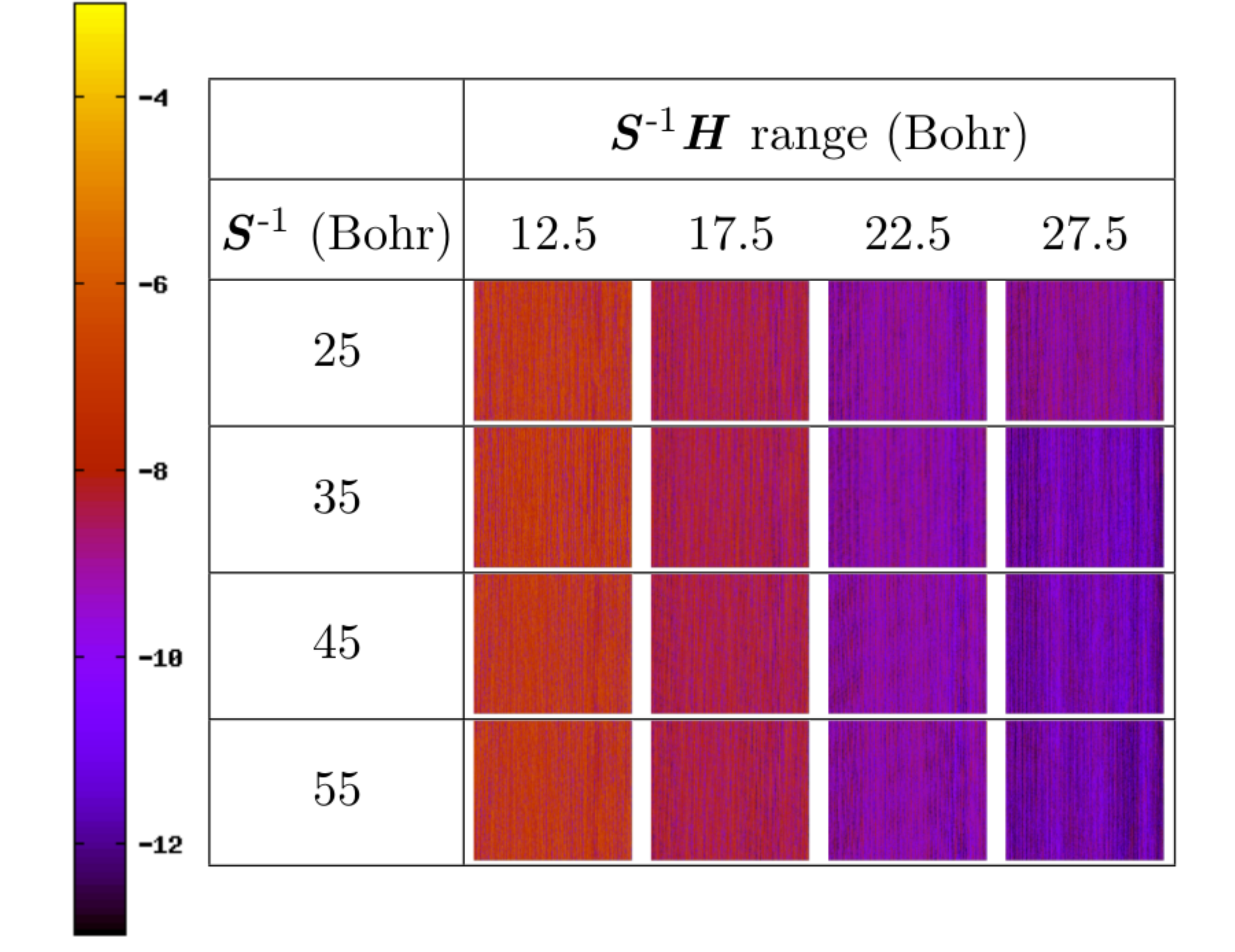}
    \end{minipage}
    \caption{Plot of the absolute values of the matrix $\textbf{\textit{U}}^{\dagger}\textbf{\textit{U}}-\textbf{\textit{I}}$ (on a base 10 log scale), illustrating the propagator unitarity for differing truncation ranges of the
 \textbf{\textit{S}}\superscript{-1} and  \textbf{\textit{S}}\superscript{-1}\textbf{\textit{H}} matrices for the C\subscript{47}H\subscript{96} molecule}
    \label{alk_unit}
   \end{center}
\end{figure}

As the range of the matrices increases the error caused by the
truncation converges towards zero, as we expect.
The \textbf{\textit{S}}\superscript{-1} matrix
converges less quickly than the
\textbf{\textit{S}}\superscript{-1}\textbf{\textit{H}} matrix, indicating that
it is more dense than the \textbf{\textit{S}}\superscript{-1}\textbf{\textit{H}} matrix.
The effect truncation of these matrices has on the unitarity of the
propagators can be seen in figure \ref{alk_unit}. We see that
the unitarity converges as the \textbf{\textit{S}}\superscript{-1}\textbf{\textit{H}}
range increases, and the propagators are converged with a range
of around $\sim$ 22.5-27.5 Bohr.
This indicates that the \textbf{\textit{S}}\superscript{-1}\textbf{\textit{H}} matrix
is indeed sparse, while the \textbf{\textit{S}}\superscript{-1} matrix is less so,
 and we can safely truncate it. It is important to not that we don't explicitly use the 
\textbf{\textit{S}}\superscript{-1} in our propagators, only the  \textbf{\textit{S}}\superscript{-1}\textbf{\textit{H}} matrix.
Although it makes sense to truncate the \textbf{\textit{S}}\superscript{-1}
matrix, given that we are truncating
\textbf{\textit{S}}\superscript{-1}\textbf{\textit{H}} and that the
Hamiltonian matrix is sparse.
We can see this by noting that the unitarity of the propagator
in figure \ref{alk_unit} is also well converged for each of the truncation ranges imposed on
the inverse overlap.

\begin{figure}[!]
  \begin{center}
%   \begin{tabular}{c c}
     \begin{minipage}[h]{0.45\textwidth}
     \includegraphics[angle=-90,origin=c,trim = 80mm 5mm 45mm 60mm, clip, width=1.\textwidth]{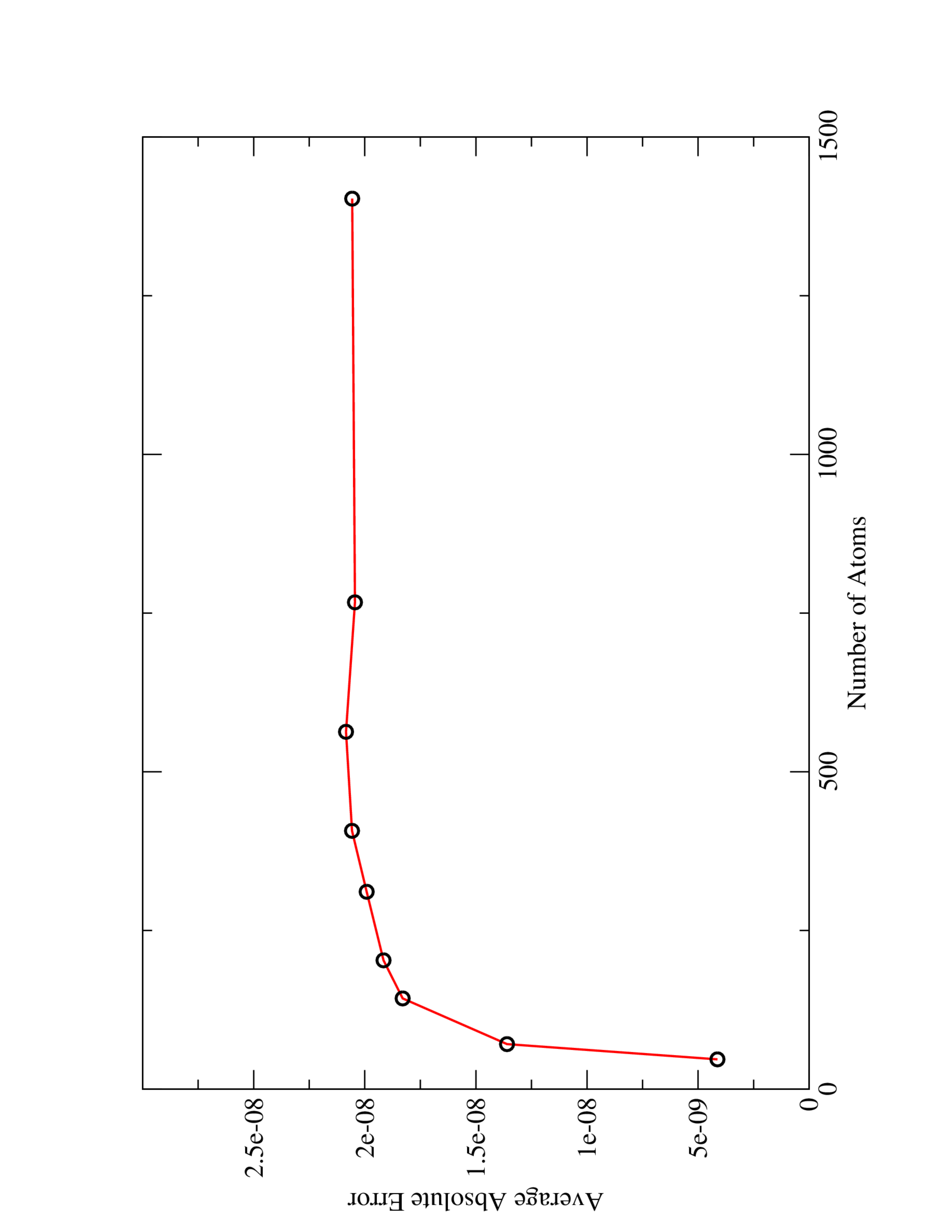}
      \end{minipage}
%    \end{tabular}
     \caption{Average absolute error in the \textbf{\textit{S}}\superscript{-1}\textbf{\textit{H}} matrix elements with system size.}
    \label{th_size}
   \end{center}
\end{figure}

As additional atoms are added the Hamiltonian matrix, overlap matrix, and the inverse overlap will vary. Increasing the system size may
therefore affect the ranges of these matrices. While we only use the \textbf{\textit{S}}\superscript{-1}\textbf{\textit{H}} matrix in our calculation, comparison of the density of both matrices have been included.
We have tested this effect by
fixing the \textbf{\textit{S}}\superscript{-1}
and \textbf{\textit{S}}\superscript{-1}\textbf{\textit{H}} ranges at 30 and 35
Bohr respectively, and examined the error in the
truncated \textbf{\textit{S}}\superscript{-1}\textbf{\textit{H}} matrix with
system size with the results shown in figure \ref{th_size}.
We see that the error changes slightly on increasing system
size, but converges as the size increases.
Consequently the propagator unitarity was found to exhibit the
same trend. This illustrates that the
\textbf{\textit{S}}\superscript{-1}\textbf{\textit{H}} is well ranged, irrespective
of system size, allowing us to impose a cut-off radii on both
of these matrices. In effect this ensures that as the system size
increases, the computational load can be made to scale linearly.

\begin{figure}[!]
  \begin{center}
%   \begin{tabular}{c c}
     \begin{minipage}[h]{0.45\textwidth}
     \includegraphics[angle=-90,origin=c,trim = 80mm 5mm 45mm 60mm, clip, width=1.\textwidth]{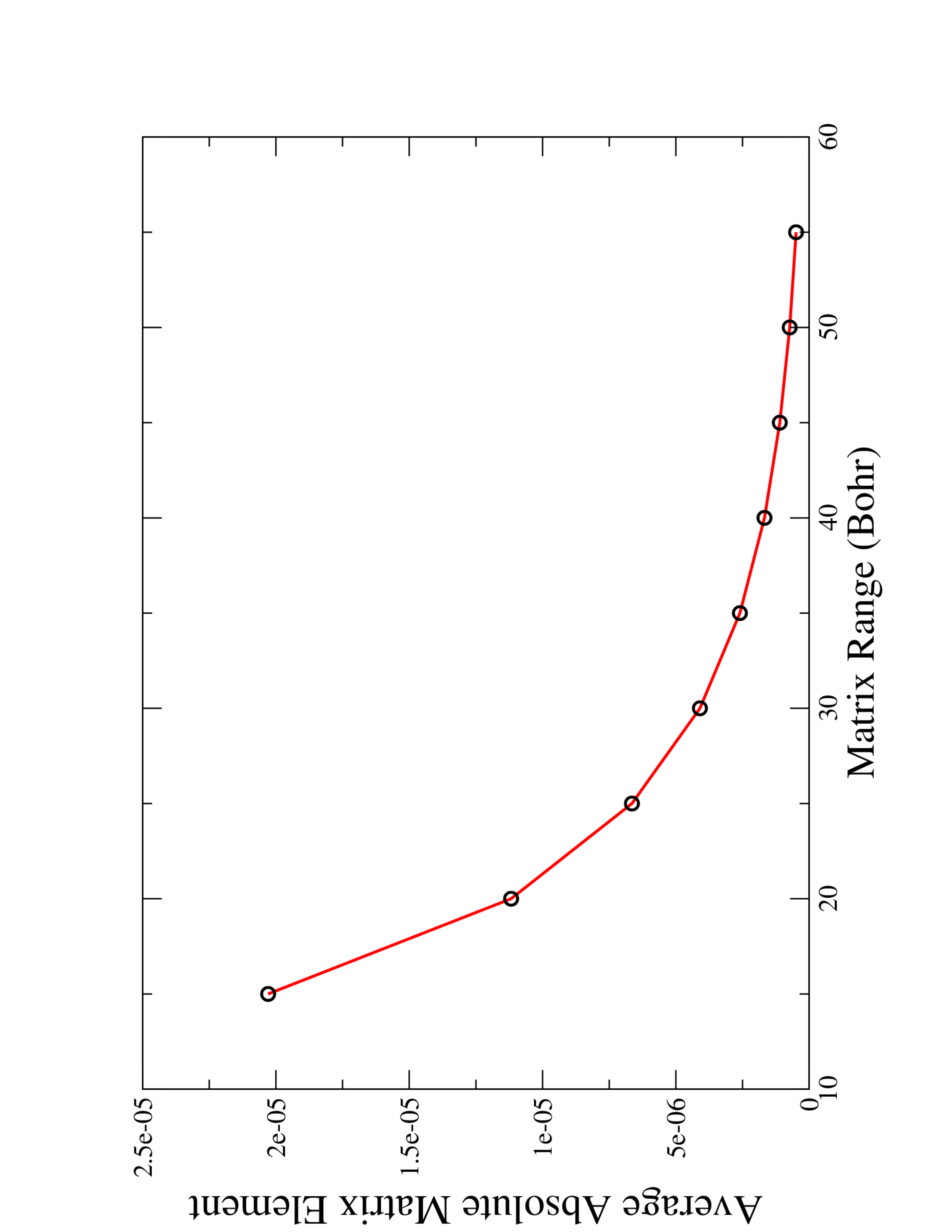}
      \end{minipage}
%    \end{tabular}
     \caption{Average value of the $\textbf{\textit{U}}^{\dagger}\textbf{\textit{U}}-\textbf{\textit{I}}$ matrix with \textbf{\textit{S}}\superscript{-1}\textbf{\textit{H}} matrix range for the C\subscript{103}H\subscript{208} molcule calculated with a SZ2P basis set.}
    \label{311_unit}
   \end{center}
\end{figure}

Similarly, increasing the number of basis functions will directly
affect the overlap matrix, and consequently the inverse overlap
and the propagator. In order to gauge the extent of this effect we
have examined the C\subscript{103}H\subscript{208} molecule with a larger basis set (SZ2P as opposed to SZP).
Exhibited in figure \ref{311_unit} is the absolute value of the $\textbf{\textit{U}}^{\dagger}\textbf{\textit{U}}-\textbf{\textit{I}}$ matrix with  \textbf{\textit{S}}\superscript{-1}\textbf{\textit{H}} matrix truncation range.
Despite the larger number of basis set functions we see that
the \textbf{\textit{S}}\superscript{-1}\textbf{\textit{H}} matrix
is still well ranged, although the range is wider when compared
to the SZP results of figure \ref{alk_unit}, and again a truncation will lead to a computational
load that scales linearly with system size.

A further point to note is that it is possible to avoid the use of the
inverse overlap matrix in the TDDFT propagation altogether.
Yam et al. have employed a Cholesky orthogonalisation scheme to
bypass the need for the inverse overlap\cite{chen_alkane}.n
However using this scheme requires the inverse
of the Cholesky decomposition, and it is not apparent that it will
be more sparse than the inverse overlap.
It is possible that this scheme might improve the calculation
of the propagator, as the orthogonalised
Hamiltonian may be more localised than our \textbf{\textit{S}}\superscript{-1}\textbf{\textit{H}} matrix. Calculating the Cholesky decomposition can be
made to scale linearly, and implementation of this
alternative method has already begun in order to contrast the two
approaches. However the parallelisation of Cholesky inversion 
is difficult given the Conquest matrix storage, and inversion of the overlap matrix remains
important.

\subsection{Density Matrix Truncation, Scaling and Limits}

\begin{figure}[!]
%  \begin{center}
     \begin{minipage}[h]{0.5\textwidth}
      \includegraphics[angle=-90,trim = 20mm 100mm 10mm 150mm, clip, width=1.\textwidth]{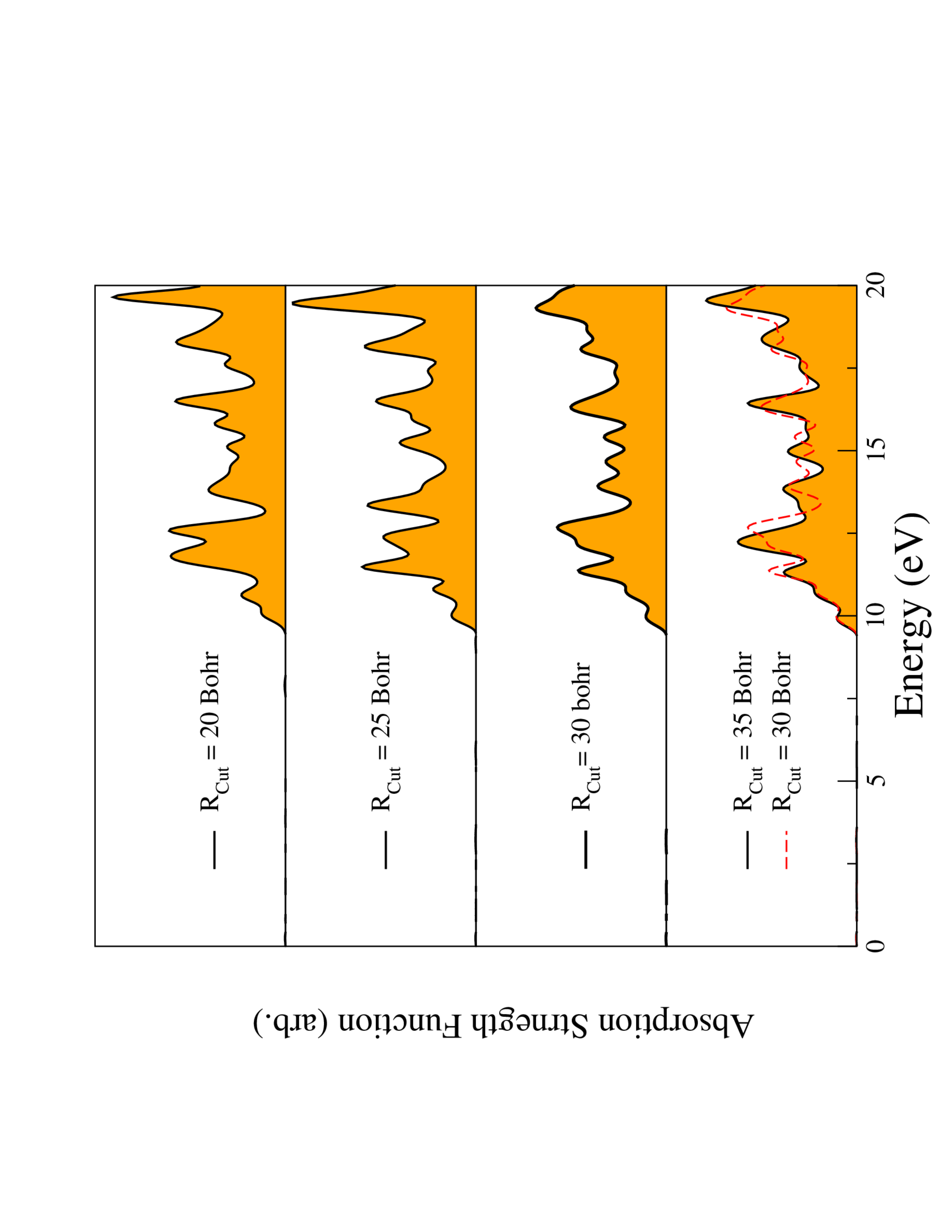}
       \end{minipage}
     \caption{K matrix truncation radii dependence: Spectra generated for
the C\subscript{47}H\subscript{96} molecule at varying density matrix
cut-off radii. (Total run time of 400 a.u. at a time step of 0.05 a.u.) }
    \label{DM_cut}
%   \end{center}
\end{figure}

Finally we examine the effect of truncating the density matrix, and
have performed calculations generating spectra for
 the C\subscript{47}H\subscript{96}
molecule at varying truncation radii, R\subscript{Cut}, of the density
 matrix. Typically for ground state calculations a
suitable typical density matrix truncation range is around
16-20 Bohr. The results can be seen in figure \ref{DM_cut},
and generally we find that as the density matrix cut-off increases
the spectra tend to converge, as expected, with higher
lying states requiring a larger cut-off to converge. We can see from the
comparison of $R\subscript{Cut}=30$ and $R\subscript{Cut}=35$ that
there is good agreement for the initial transitions, as well as the
general shape of the spectra.

\begin{figure}[h]
  \begin{center}
     \begin{minipage}[h]{0.5\textwidth}
      \includegraphics[trim = 20mm 20mm 70mm 20mm, clip, width=1.\textwidth]{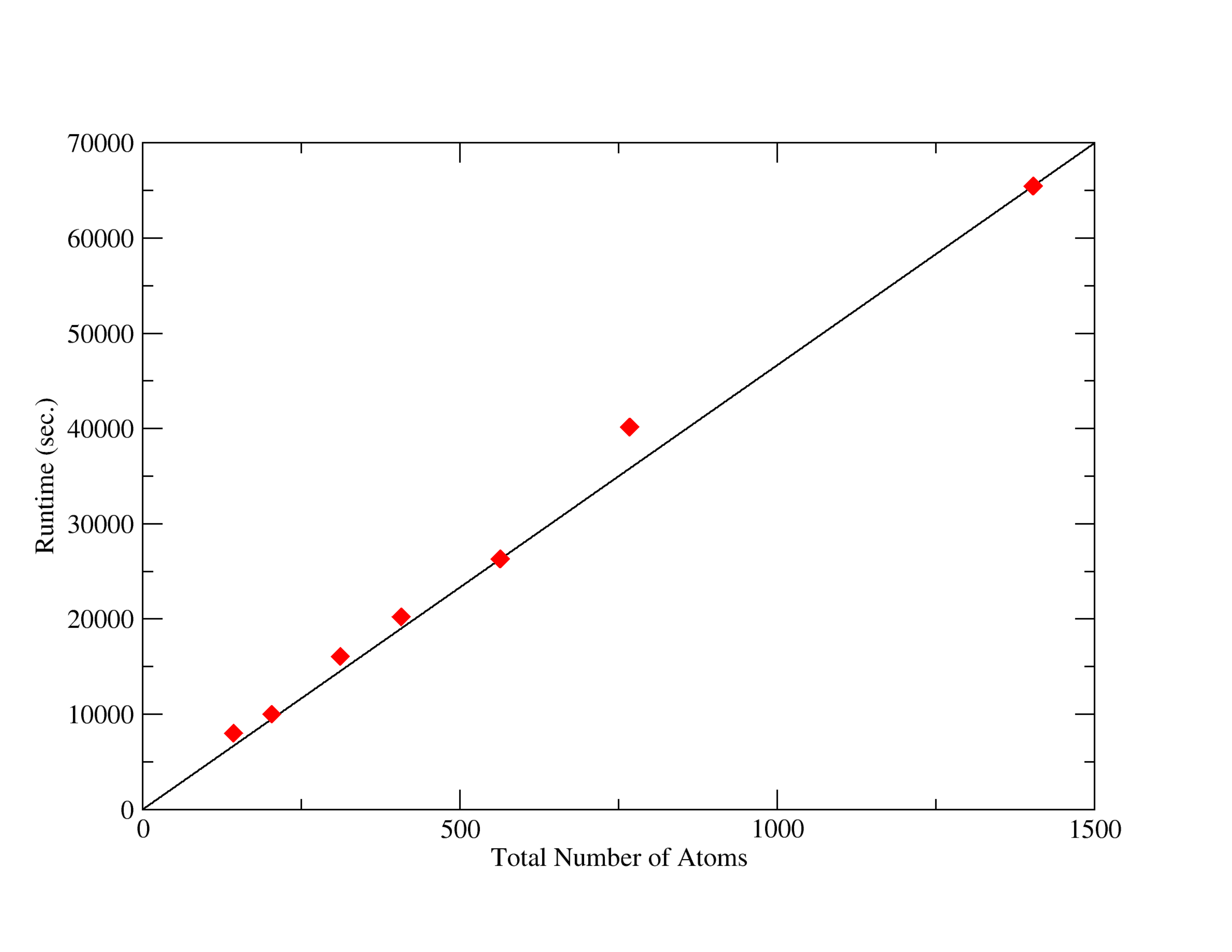}
       \end{minipage}
     \caption{Computational TDDFT run time versus system size for long chain alkane molecules. The system was run with a timestep of 0.05 a.u. for a total time of 10 a.u. A matrix truncation range, $R\subscript{Cut}=35$ a.u., has been applied. }
    \label{ON_time}
   \end{center}
\end{figure}

Applying this $R_{Cut}=35$ Bohr cut-off (along with a cut off of 35 Bohr.
on the $\textbf{\textit{S}}^{-1}\textbf{\textit{H}}$ matrix) we can examine the
computational scaling with system size, with the results exhibited
in figure \ref{ON_time}. Clear linear scaling of the computational workload 
up to well over 1000 atoms is exhibited, illustrating the potential power of the method.

Finally a few comments on the limits of the approach must be made. TDDFT for long-range charge transfer is well known to be poorly described by local and semi-local 
functionals\cite{NON_LOCAL_TDDFT}. While we have employed LDA and GGA functionals here, linear scaling
exact exchange has also been recently implemented in the Conquest code, allowing the use of
non-local functionals with this approach in the future.

While the near-sightedness principle 
dictates that the ground-state density matrix is exponentially localised for well gapped systems, 
there is no formal justification for the localisation of the response density matrix. 
As noted in \cite{onetep_linear}, for systems with well localised excitations it would be expected that 
the response density matrix could be truncated safely and linear scaling achieved, while for systems 
with delocalised excitations this will not be the case.

\section{Conclusions}

We have outlined our implementation of real-time time dependent
density
functional theory in the Conquest $\mathcal{O}$(N) code.
We have demonstrated the soundness of the implementation through
benchmark tests for small molecules, and also discussed the effect of
basis set and system sizes on the results.

$\mathcal{O}(N)$ approaches utilise the density matrix, as opposed to
working directly with Kohn-Sham orbitals, providing a route
to the linear scaling computational time with system size by
its truncation. We have discussed the range of our propagator
matrices for an alkane chain test system, and the implications
of this matrix truncation on the unitarity of the propagation.
Similarly we have examined the effect of truncating the density
matrix on the calculated optical absorption spectra, showing that
the range required is much more extended than that required for
converged ground state properties.
Nevertheless, we have shown that accurate linear scaling
TDDFT calculations are practical. While the
impact of localisation cut-off in the charge density
matrix on these TDDFT calculations is a topic warranting further study, we
have shown that in truncating these matrices at a suitable point
we obtain a computational load that increases linearly with system
size. This offers a complementary approach to the usual Casida linear
response approach: linear response TDDFT is well suited to relatively
small systems, while linear scaling RT-TDDFT offers a viable method
for studying excitations in large systems. We have shown linear
scaling beyond 1,000 atoms, and 10,000+ atoms are perfectly
practical with the excellent parallel scaling available in Conquest.

\section*{Acknowledgements}
C.O'R. is supported by the MANA-WPI project and D.R.B. was
funded by the Royal Society.
We thank Umberto Terranova for useful
discussions. This work made use of the facilities of ARCHER, the UK's national high-performance computing service, which is provided by UoE HPCx Ltd at the University of Edinburgh, Cray Inc and NAG Ltd, and funded by the Office of Science and Technology through EPSRC's High End Computing Programme. Calculations were performed at HECToR through the UKCP Consortium. The authors acknowledge the use of the UCL Legion High Performance Computing Facility, and associated support services, in the completion of this work.

\bibliography{TDDFT}

\end{document}